\documentclass[11pt,aps,preprint,tightenlines,nofootinbib]{revtex4}
\usepackage{epsfig}
\usepackage{graphics}
\usepackage{bm}
\usepackage{color}

\newcommand{\ff}{\bm{f}}

\newcommand{\nab}{\mbox{\boldmath $\nabla$} {}}

\newcommand{\bra}[1]{\langle #1\rangle}

\def\Pm{\mbox{\rm Pr}_{\rm M}}

\def\Rey{\mbox{\rm Re}}

\newcommand{\EQ}{\begin{equation}}
\newcommand{\EN}{\end{equation}}

\newcommand{\DD}{{\rm{D}}}

\usepackage{times}

\begin{document}
\title
{The evolution of the primordial magnetic field since its generation}
\preprint{NORDITA-2015-83}

\author{Tina Kahniashvili}
\email{tinatin@andrew.cmu.edu} \affiliation{The McWilliams Center
for Cosmology and Department of Physics, Carnegie Mellon University,
5000 Forbes Ave, Pittsburgh, PA 15213, USA} \affiliation{Department
of Physics, Laurentian University, Ramsey Lake Road, Sudbury, ON P3E
2C, Canada} \affiliation{Abastumani Astrophysical Observatory, Ilia
State University, 3-5 Cholokashvili Avenue, Tbilisi, 0162, Georgia}

\author{Axel Brandenburg}
\email{Axel.Brandenburg@Colorado.edu}
\affiliation{Laboratory for Atmospheric and Space Physics, University of Colorado, Boulder, CO 80303, USA}
\affiliation{JILA and Department of Astrophysical and Planetary Sciences, University of Colorado, Boulder, CO 80303, USA}
\affiliation{Nordita, KTH Royal Institute of Technology and Stockholm University,
Roslagstullsbacken 23, 10691 Stockholm, Sweden}
\affiliation{Department of Astronomy, AlbaNova University Center,
Stockholm University, 10691 Stockholm, Sweden}

\author{Alexander G.\ Tevzadze}
\email{aleko@tevza.org} \affiliation{Faculty of Exact and Natural
Sciences, Tbilisi State University, 3 Chavchavadze Avenue, Tbilisi,
0179, Georgia} \affiliation{Abastumani Astrophysical Observatory,
Ilia State University, 3-5 Cholokashvili Avenue, Tbilisi, 0162,
Georgia}

\begin{abstract}

We study the evolution of primordial magnetic fields in an expanding cosmic
plasma. For this purpose we present a comprehensive theoretical model to
consider the evolution of MHD turbulence that can be used over a wide range
of physical conditions, including cosmological and astrophysical
applications. We model different types of decaying cosmic MHD turbulence in
the expanding universe and characterize the large-scale magnetic fields in
such a medium. Direct numerical simulations of freely decaying MHD turbulence
are performed for different magnetogenesis scenarios: magnetic fields
generated during cosmic inflation as well as electroweak and QCD phase
transitions in the early universe.

Magnetic fields and fluid motions are strongly coupled due to
the high Reynolds number in the early universe.
Hence, we abandon the simple adiabatic dilution model to estimate magnetic
field amplitudes in the expanding universe and include turbulent mixing
effects on the large-scale magnetic field evolution. Numerical simulations
have been carried out for non-helical and helical magnetic field
configurations. The numerical results show the possibility of inverse
transfer of energy in magnetically dominated non-helical MHD turbulence. On
the other hand, decay properties of helical turbulence depend on whether the
turbulent magnetic field is in a weakly or a fully helical state.

Our results show that primordial magnetic fields can be considered as
a seed for the observed large-scale magnetic fields in galaxies and clusters.
Bounds on the magnetic field strength are obtained and are consistent with
the upper and lower limits set by observations of extragalactic magnetic
fields.

\end{abstract}

\maketitle

\section{Introduction}

Understanding the origin and evolution of cosmic magnetism is one of the
challenging questions of modern astrophysics. The major questions include
theoretical as well as observational aspects of the problem: when and how was
the cosmic magnetic field generated? How did it evolve during the expansion
of the universe? What are modern observational constraints on the magnetic
fields at large scales? Are magnetic fields observed at galactic and
extragalactic scales of cosmological or astrophysical origin? The types of
turbulence considered here are characterized by a strong random initial
magnetic field. The interaction with the velocity field leads too inverse
spectral transfer toward large scales that is unknown in non-magnetic
turbulence.\footnote{The paper is based on the presentation by Tina
Kahniashvili {\it Cosmic Magnetic Fields: Origin, Evolution, and Signatures}
at the Turbulent Mixing and Beyond Workshop 2014 {\it 'Mixing in Rapidly
Changing Environments - Probing Matter at the Extremes'}.}

The goal is to identify important properties of cosmic magnetic turbulence in
the expanding universe. Properties of decaying MHD turbulence in primordial
plasma link magnetogenesis scenarios operating in the early universe with the
constraints on the large-scale magnetic fields set by present observations.
Hence, studying the magnetic field evolution, we can identify likely
magnetogenesis scenarios responsible for exciting seed fields in the early
universe and exclude unlikely ones using constraints set by modern or future
observations.

The problem of cosmological magnetogenesis is guided by recent observations
of large-scale magnetic fields.
Indeed, galaxies are known to
have magnetic fields that are partly coherent on the scale of the galaxy with
field strengths reaching $10^{-6}$ Gauss (G) (see Refs.\
\cite{Widrow:2002ud,Vallee,Giovannini:2006kg,Beck:2011gv,
VanEck:2010ka,Fletcher:2010wt} and references therein).  These magnetic
fields are the result of amplification of initial weak seed fields of unknown
nature. Moreover, it is now clear that $\mu$G-strength magnetic fields were
already present in normal galaxies (like our Milky Way) when the universe was
less  than half of its present age
\cite{Bernet:2008qp,Kronberg:2007dy,Beck:2011he}.
This poses strong limits on
the seed magnetic field strength and its amplification timescale.

From a theoretical point of view there are two scenarios that can lead
to the generation of magnetic fields at extragalactic scales
\cite{Durrer:2013pga}: a bottom-up
(astrophysical) scenario, where the seed field is typically very weak and the
observed large-scale magnetic field is transported from local sources within
galaxies to larger scales \cite{Kulsrud:2007an}, and a top-down
(cosmological) scenario where a significant seed field is generated prior to
galaxy formation in the early universe on scales that are large at the present time
\cite{Kandus:2010nw}. The major theme of this review is to discuss the
evolution, structure, and effects of cosmic magnetic fields with the goal
to better understand its origin and observational signatures.

We will briefly discuss cosmic magnetohydrodynamic (MHD) turbulence in order
to understand the magnetic field evolution. MHD turbulence in the context of
astrophysical plasma processes has been studied for a long time. On the other
hand, the effects of MHD turbulence in cosmological contexts has received
attention only in recent years \cite{b1}. Simulations show that the kinetic
energy of turbulent motions in galaxy clusters can be as large as 5--10\% of
the thermal energy density \cite{Kravtsov:2012zs}. This can influence the
physics of clusters \cite{Subramanian:2005hf}, and at least should be modeled
correctly when performing large-scale simulations
\cite{Vazza:2006cp,Fang:2008ad,Greif:2008qqa,Pakmor:2011ht,Cho:2014oca,Vazza:2014jga}.
Turbulent motions can also affect cosmological phase transitions; see
Refs.~\cite{Espinosa:2010hh,Espinosa:2007,Medina:2014} and references
therein). Turbulence can be generated by a small initial cosmological
magnetic fields. Understanding mechanisms for exciting primordial turbulence
is an important goal. We argue that even if the total energy density present
in turbulence is small, its effects might be substantial because of the
strongly nonlinear nature of the relevant physical processes.

Recent important observations
Refs.~\cite{Neronov:1900zz,Tavecchio:2010ja,Essey:2010nd,Taylor:2011bn,Huan:2011kp,
Vovk:2011aa,Dolag:2010ni,Takahashi:2011ac,Dermer:2010mm}
(also Ref.~\cite{Finke:2015ona} for recent study, and Ref.~\cite{Arlen:2012iy}
for discussions on possible uncertainties in the measurements of blazar
spectra), suggest the existence
of magnetic fields in the universe at scales large enough to suggest a primordial origin \cite{Durrer:2013pga}.
This result is robust to potential plasma instabilities of the two-stream family
\cite{Miniati:2012ge,Broderick:2011ab,Lemoine:2014gca}.
Prior
to these observations, there existed only {\it upper} limits of the order of a few
nG for the intergalactic magnetic field. These were obtained through Faraday
rotation of the cosmic microwave background (CMB) polarization plane
\cite{Kosowsky:1996yc,Harari:1996ac,Giovannini:2008aa,Campanelli:2004pm,Scoccola:2004ke,
Kahniashvili:2008hx,Kosowsky:2004zh,Pogosian:2011qv,Pshirkov:2015tua,Kahniashvili:2010wm,Ade:2015cao}
and Faraday rotation of polarized emission of distant quasars
\cite{a1,a11,a2,a3}. Other tests to derive {\it upper} limits on large-scale
correlated magnetic fields are based on their effect on the CMB (see
Ref.~\cite{Ade:2015cva} and references therein), \cite{Adams:1996cq,Subramanian:1998fn,Subramanian:2002nh,Subramanian:2003sh,
Durrer:1999bk,Mack:2001gc,Pogosian:2001np,Caprini:2003vc,Lewis:2004ef,Kahniashvili:2005xe,Kahniashvili:2005vu,
Kahniashvili:2006hy,Kristiansen:2008tx,Paoletti:2008ck,
Shaw:2009nf,Paoletti:2010rx,Kunze:2010ys, Yamazaki:2010jw,
Ichiki:2011ah,Yamazaki:2011eu,Kunze:2011bp, Trivedi:2011vt,
Paoletti:2012bb,Trivedi:2013wqa,Chen:2013gva,Yamazaki:2013hda,
Ballardini:2014jta,Kahniashvili:2014dfa}, CMB distortions
\cite{Jedamzik:1999bm,Dent:2012ne,Jedamzik:2013gua,Miyamoto:2013oua,Amin:2014ada,
Chluba:2015lpa,Wagstaff:2015jaa,
Kunze:2013uja,Kunze:2014eka,Tashiro:2013yea}, the broken isotropy limits,
\cite{Bernui:2008,Durrer:1998ya,Chen:2004nf,Samal:2009,Demianski:2007,Brown:2005,
Kahniashvili:2008sh,Naselsky:2008ei,Kim:2009gi,Ade:2013nlj}, big bang
nucleosynthesis (BBN) data
\cite{Kawasaki:2012va,Yamazaki:2012jd,Yamazaki:2014fja}, or large scale structure (LSS)
formation \cite{Tashiro:2006uv,Tashiro:2005hc,Tashiro:2005ua,Jedamzik:1996wp,
Tashiro:2009hx,Shaw:2010ea,Yamazaki:2010nf,Widrow:2011hs,Schleicher:2008hc,
Schleicher:2011jj,Fedeli:2012rr,Kahniashvili:2012dy,Sethi:2004pe,Pandey:2012ss,
Pandey:2014vga,magfields2,magfields3,Schleicher:2008aa,
Sethi:2009dd,Sethi:2008eq,Venumadhav:2014tqa,Ade:2015kva,Vasiliev:2014vpa}. The {\it lower
limit} on the intergalactic magnetic field in voids of order $10^{-18}$ G on
1\,Mpc scales is a puzzle of modern astrophysics (see Ref. \cite{Miniati:2010ne}),
and could very well be the result of the amplification of a primordial
cosmological field \cite{Dolag:2010ni}.

In what follows we review recent efforts which include the pioneering studies
of primordial magnetic field evolution through cosmological phase
transitions; see Refs.~\cite{Kahniashvili:2009qi,Kahniashvili:2010gp,
Kahniashvili:2012vt,Tevzadze:2012kk,Kahniashvili:2012uj,Brandenburg:2014mwa,kbtv15,bkt15,kbdty15,
Brandenburg:2016odr,Kahniashvili:2015msa}.
The decay of cosmic magnetic field in the universe has been analyzed through numerical
simulations of decaying MHD turbulence. Major findings include (i) the
possibility of the inverse transfer of {\it non-helical} causally generated
magnetic fields \cite{Brandenburg:2014mwa}; (ii) fast growth of vorticity in
the magnetized universe \cite{Kahniashvili:2012vt}; (iii) growth of helical
structures at large scales for partially helical magnetic fields generated at
cosmological phase-transitions \cite{Tevzadze:2012kk,Kahniashvili:2012uj,kbtv15,Kahniashvili:2015msa},
and more interestingly the absence of the inverse cascade for
inflation-generated fully or partially helical magnetic fields
\cite{bkt15,Brandenburg:2016odr}.

\section{Modeling MHD Turbulence in the Universe}

The origin of the cosmic magnetic field has been discussed for decades,
starting with Enrico Fermi's paper of 1949 \cite{fermi}. The approach presented
below is novel in several ways. (i) Primordial magnetic fields are generally
analyzed in the ``frozen-in'' approximation due to a high conductivity of
cosmic plasma, when the magnetic field evolves only due to the dilution
of field lines as the universe expands. In contrast, we account for the
actual coupling between the magnetic field and the cosmic plasma, which
leads to major differences with the frozen-in approximation at some epochs.
(ii) Much work on MHD turbulence is focused on specific astrophysical
objects (such as galaxies, clusters, interstellar medium, or stellar
magnetosphere). Instead we have developed a comprehensive theoretical
framework to consider the evolution of MHD turbulence over a wide range of
physical conditions, beyond any specific application. (iii) Cosmic MHD
turbulence is usually studied within one of two limiting cases, the viscous
(optically thick) or free-streaming (optically thin) regimes. These two
regimes differ in the form of viscous or drag forces. Realistic turbulent behavior is
somewhere in between these two limits, and the numerical simulations
have the capability to describe adequately a smooth transition between these
two regimes.

As noted above, several astrophysical observations show the presence of a
large-scale correlated magnetic fields in the universe. The recent study by
Dolag et al.\ \cite{Dolag:2010ni} concludes that these magnetic fields are
most likely seeded by a field of primordial origin. In fact, many different
mechanisms of cosmological seed magnetic field generation have been proposed.
Some of these employ symmetry breaking during phase transitions (e.g.\
electroweak or QCD) \cite{phase,Quashnock:1988vs,t1,t3,t4,t5,t6,t7,t8,t9,t10,t11,t12,t13,t14,t15,t16,t17,t18,Grasso:1997nx,Stevens:2012zz}.
On the other hand, if the magnetic field originated during a cosmological
phase transition, its configuration is strongly limited by causality
\cite{hogan}: the correlation length of the magnetic field cannot exceed the
Hubble horizon at the moment of field generation. The causality condition
combined with the divergence-free field condition implies a magnetic
energy spectrum at large scales $E_M(k) \propto k^4$ \cite{causal} (the
so-called Batchelor spectrum \cite{dav}). Recent numerical simulations
\cite{Tevzadze:2012kk,Kahniashvili:2012vt,Kahniashvili:2010gp,Brandenburg:2014mwa,Kahniashvili:2012uj}
confirm that cosmological turbulence produces a Batchelor spectrum completely
independently of initial conditions present in the cosmic plasma. Combining
this {\it causal} spectrum with the requirement that the total energy density
of the magnetic field be less than $10\%$ of the radiation energy density (to
be consistent with standard BBN) leads to a strong limit on the smoothed
amplitude of the magnetic field at large scales of the order of $10^{-26}$
to $10^{-19}$ Gauss at 1\,Mpc \cite{Caprini:2001nb}, although the effective
value of the magnetic field derived through its total energy density is high
enough, of the order of $10^{-6}$ Gauss \cite{Kahniashvili:2009qi}. (Note
that this argument does not account for further evolution of the magnetic
field in MHD turbulence). Taking into account that the magnetic field effects
are mostly determined by effective values (i.e.\ the total energy density),
and noticing that the extremely low limits at large scales of causal fields
are consequences of normalization (smoothing procedure), the upper bounds
have been re-determined in terms of the effective strength of the
magnetic field; see Refs.\
\cite{Kahniashvili:2008hx,Kahniashvili:2010wm,Kahniashvili:2012dy}.
The BBN limits have also been re-analyzed by accounting for the MHD evolution
of {\it toy} magnetic fields throughout expansion of the universe
\cite{Kahniashvili:2012uj}.

Our particular interest lies in helical magnetic fields that
can be generated in the early universe; see Refs.\
\cite{Cornwall:1997ms,Giovannini:1997eg,Field:1998hi,Vachaspati:2001nb,Tashiro:2012mf,
Sigl:2002kt,Subramanian:2004uf,Campanelli:2005ye,
Semikoz:2004rr,DiazGil:2007dy,Campanelli:2008kh,Campanelli:2013mea,Jain:2012jy}
and references therein. There are two main motivations for considering helical
seed magnetic fields: (i) the presence of helical magnetic fields in the
early universe can be related to the lepto- and baryogenesis problems
\cite{Long:2013tha}; (ii) it sheds light on the evolution of helical magnetic fields in
stellar magnetospheres, AGNs, and
voids \cite{Tashiro:2013ita,Tashiro:2014gfa}.

An exception to the Batchelor spectrum (spectral index $n=4$) is the
possibility of inflationary magnetogenesis, in which the spectral index of
the magnetic field could be less than $+1$, and the simplest option is a
scale-invariant spectrum with $n \rightarrow -1$
\cite{inflation,ratra,i1,i2,i3,i4,i5,i6,i7,i8,i9,Bamba:2012mi,Jimenez:2011uia,
Membiela:2010rv,Membiela:2012ju,Fujita:2012rb,
Motta:2012rn,Bonvin:2011dt,Byrnes:2011aa,Das:2010ywa,Durrer:2010mq,
Jimenez:2010hu,Bamba:2011si,Martin:2007ue,Caldwell:2011ra}. Inflation-generated magnetic field scenarios should be
considered with some caution due to the possibility of significant
backreaction \cite{Demozzi:2009fu,Kanno:2009ei,Urban:2011bu,b22}, which is
not an issue for the phenomenological, effective classical model; see Ref.\ \cite{Tasinato:2014fia} and references therein. The first simulations
describing the inflation-generated magnetic field coupled to the primordial
plasma suggested that the presence of an initial magnetic field
leads to large-scale turbulent motions in the rest plasma
\cite{Kahniashvili:2012vt}. Ongoing research consists in the study of
inflation-generated helical magnetic field (with a scale-invariant $k^{-1}$
spectrum) evolution during the expansion of the universe \cite{kbdty15,bkt15,Brandenburg:2016odr}.
Simulations show that inflation-generated magnetic fields retain
information about initial conditions. In other words, they decay very slowly
when compared with phase transition-generated fields. Magnetic fields are almost
``frozen-in'' the primordial plasma at large scales, where causality allows
interaction only at scales smaller than the Hubble horizon and they
correspondingly retain their initial spectral shape. On the other hand,
within the causal horizon, the magnetic seed field interacts with cosmic plasma
leading to the excitation of kinetic motions (turbulent velocities); see below.
We also discuss an alternative approach where cosmic magnetic fields
originate during the late stages of the evolution of the universe
\cite{Widrow:2011hs,Brandenburg:2013vya}.
In this case the correlation length
is strongly limited by the causality requirement. Due to the sharp
spectral shape at large scales, the magnetic field amplitude might be low. The
simplest astrophysical magnetogenesis mechanism invokes the ejection of magnetic flux from
compact systems such as AGNs or supernovae
\cite{Colgate,Xu:2009if}. In this scenario the generation of a strong
magnetic field is ensured by its extremely fast generation due to rapid
rotation of the object \cite{Kulsrud:2007an}. Other mechanisms are based on
the generation of a small seed by plasma processes
\cite{nature2012,Alves:2011pp,Miniati:2010ne}, which are then amplified by
MHD dynamo mechanisms \cite{amplification,am1}.

\section{The model}

As mentioned above, we focus on the magnetic field evolution during the
expansion of the universe from the moment of magnetic field generation until
today. Over this lengthy period, the magnetic field is affected by different
physical processes that result in amplification as well as damping: the
complexities of the problem are due to the strong coupling between magnetic
field and turbulent motions. First, to account for the cosmological expansion
we must reformulate the MHD equations in terms of comoving quantities
\cite{beo96}. Specific epochs most relevant to the final configuration of the
primordial magnetic field are related to cosmological phase transitions,
neutrino decoupling, nucleosynthesis, recombination, and reionization; see
\cite{axel-mode,Durrer:2013pga} for reviews and
Refs.~\cite{robi,Jedamzik:1996wp,Subramanian:1997gi,
Dimopoulos:1996nq,banerjee,Wagstaff:2013yna,kulsrud}.

In the following, we discuss numerical simulations performed with the {\sc
Pencil Code}. This public MHD code (\url{https://github.com/pencil-code})
(see also Ref.~\cite{db}) is particularly well suited for simulating
turbulence owing to its high spatial (sixth order) and temporal (third order)
accuracy, while still taking advantage of the finite difference in terms of
speed and straightforward parallelization.
Recent results from the {\sc Pencil Code} include MHD turbulence simulations at the electroweak
or QCD phase transitions
\cite{Kahniashvili:2010gp,Kahniashvili:2012vt,Tevzadze:2012kk,Kahniashvili:2012uj,Brandenburg:2014mwa}.

\paragraph{Numerical Technique.}
By default, the {\sc Pencil Code}
{solves the MHD equations for the logarithmic
density $\ln\rho$, flow velocity $\bm{v}$,}
and the magnetic vector potential $\bm{A}$ as follows:
\begin{eqnarray}
\frac{\DD}{\DD\eta} \ln\rho &=&- \bm{\nabla}\cdot\bm{v} ~,\\
\frac{\DD}{\DD\eta} \bm{v} &=&\bm{J}\times\bm{B}-c_s^2\bm{\nabla}\ln\rho
+\bm{f}_{\rm visc} ~,
\\
\frac{\partial}{\partial \eta} \bm{A} &=&
\bm{v}\times\bm{B}
+{\bm f}_M+\lambda \nabla^2{\bm A} ~.
\label{mhd3}
\end{eqnarray}
{Here, $\eta$ is the conformal time and $\DD/\DD\eta \equiv
\partial/\partial\eta+\bm{v}\cdot\bm{\nabla}$ is the advective time
derivative},
$\bm{f}_{\rm visc}=\nu\left(\nabla^2{\bm v}
+{\textstyle\frac{1}{3}}\bm{\nabla}\bm{\nabla}\cdot\bm{v}+\bm{G}\right)$ is
the {compressible viscous force for constant $\nu$,}
$G_i=2{\sf S}_{ij}\nabla_j\ln\rho$,
{and}
${\sf S}_{ij}=\frac{1}{2}(v_{i,j}+v_{j,i})-\frac{1}{3}\delta_{ij}v_{k,k}$
{is the traceless rate-of-strain tensor.}
The pressure is given by $p=\rho c_s^2$, where $c_s=1/\sqrt{3}$ is the
{speed of sound in the case of an ultra-relativistic gas, and}
$\bm{J}=\bm{\nabla}\times\bm{B}/4\pi$ is
the current density.

In our simulations we use a vanishing magnetic forcing term
${\bm f}_M=\bm{0}$ everywhere,
except for the purpose of producing initial conditions, as explained below.

\paragraph{Initial Conditions.}

{To produce initial conditions, we run the simulation
for a short time
($\Delta t\approx0.5 \lambda_1/c_s$) with a random (in time) $\delta$-correlated
magnetic force $\ff_M$ in Eq.~(3).}
{The forcing term is composed of plane monochromatic waves
pointing randomly in all possible directions with an average wavenumber $k_0$}
and fractional helicity
$\bra{\ff_M\cdot\nab\times\ff_M}/\bra{k_0\ff_M^2}=2\sigma/(1+\sigma^2)$.
{Here $\sigma$ is the parameter characterizing the initial
forcing.}
Initial conditions for the magnetic and velocity fields
produced from such a procedure have the advantage of being turbulent, still
self-consistent solutions of the MHD equations.

\paragraph{Effective Magnetic Field Characteristics.}
{A magnetic field generated during phase transitions through
any magnetogenesis scenario should satisfy the causality condition
\cite{hogan,beo96,causal}.}
The maximal correlation length $\xi_{\rm max}$ of a causally generated
primordial magnetic field
{should be shorter then}
the Hubble radius at the time of generation, $H_\star^{-1}$. We define the
parameter $\gamma= \xi_{\rm max}/ H_\star^{-1} \leq 1$, which can
{describe the number of primordial magnetic field bubbles
inside the Hubble radius, and thus}
$N \propto \gamma^{-3}$. To account for the universe expansion we use
comoving length, {which is measured today and corresponds to
the Hubble radius at the moment of magnetic field generation.}
{Comoving length should be inversely proportional to the phase
transition temperature ($T_\star$)}:
\begin{equation}
\lambda_{H_\star} = 5.8 \times 10^{-10}~{\rm
Mpc}\left(\frac{100\,{\rm GeV}}{T_\star}\right)
\left(\frac{100}{g_\star}\right)^{{1}/{6}} ~. \label{lambda-max}
\end{equation}

{For the QCD phase transition ($g_\star=15$ and $T_\star=0.15$
GeV), the comoving length equals 0.5\,pc, while for the electroweak phase
transition ($g_\star=100$ and $T_\star =100$ GeV) it should be equal to $6
\times 10^{-4}$ pc. In all cases, the correlation length
of the primordial magnetic field
should not exceed the comoving value of the Hubble radius:}
$\xi_{\rm max} \leq \lambda_H$. Obviously,
{the latter condition accounts {\it only} for the increase of
the correlation length due the expansion of the universe, and does not account for
the effects of cosmic MHD turbulence (free decay or an inverse cascade in
the case if primordial magnetic fields have nonzero helicity).}
{Note, that the number of bubbles inside}
the Hubble radius is around 6 ($\gamma \simeq 0.15$) for the QCD phase
transition and around 100 ($\gamma \simeq 0.01$) for the electroweak phase
transition.
{Thus, the maximal correlation length for the QCD  and
electroweak phase transitions should be 0.08\,pc and $6 \times 10^{-6}$\,pc,
respectively.}
On the other hand, the correlation length is unlimited in the case of
inflation-generated magnetic fields.

As mentioned above, the primordial magnetic field contributes to the
relativistic component and thus
{the total energy density of the primordial magnetic field
$\rho_B (a_{\rm N})$, where $a_{\rm N}$ is the scale factor during
nucleosynthesis, is limited by the BBN bound: it cannot exceed $10\%$ of the
radiation energy density $\rho_{\rm rad}(a_{\rm N})$.}
{It is straightforward to see that the maximal value of the
effective magnetic field defined through the total magnetic energy does not
depends on the temperature at the moment of generation ($T_*$), and depends weakly on
the relativistic degrees of freedom ($\gamma$) at the moment of the magnetic
field generation.}

{The dominant contribution to the magnetic field energy
density comes from the given length-scale, the so-called {\it integral scale},
where the magnetic field strength reaches its maximum.}
Thus, when dealing with phase transition-generated magnetic fields, we adopt
the following idealizing approximation:
{ we generate initial conditions for freely decaying
turbulence simulations by running a numerical simulation of forced MHD
equations for a short time interval. The external electromagnetic force, intended
to generate a turbulent state, is introduced in the form of $\delta$ functions that
peak at a characteristic wavenumber, $k_0=2\pi/\xi_{0}^{-1}$. This yields
random magnetic fields with correlation length $\xi_0$. Thus, the magnetic
field strength at the characteristic length scale is $B^{\rm
(eff)}=\sqrt{8\pi \rho_B}$. The characteristic length scale of
the initial turbulent state $\xi_0$ is set by the size of the largest
magnetic eddies, because the primordial magnetic field evolves with the primordial
MHD turbulence. In this approach, the characteristic length scale of the magnetic
field is set by the bubble size of the phase transition when magnetogenesis
occurs.}

\paragraph{Magnetic Field Spectrum.}
The interaction between magnetic field and plasma gives rise to kinetic
motions, and the turbulent backreaction results in spreading of the
{spectral energy density of magnetic field over a range of
wavenumbers. At scales longer then the integral scale of the turbulence
(small wavenumbers), the spectral energy density develops into the form of a power
law $E_M = A k^{n}$, where $A$ is a normalization constant, and $n$ is the
spectral index. The spectrum of the turbulent magnetic field can be
determined by the spectral expansion}
of the two-point correlation function of the magnetic field $\langle B_i({\bf
x}) B_j({\bf x}+{\bf r}) \rangle$, whose Fourier transform with respect to
$\bm{r}$ gives the spectral function
\begin{eqnarray}
F_{ij}^M\!({\bf k}) =  \label{eq:4.1}
P_{ij}({\bf k}) \frac{E^M\!(k)}{4\pi k^2}  + i
\varepsilon_{ijl} {k_l} \frac{H^M\!(k)}{8\pi k^2}.
\end{eqnarray}
Here, $P_{ij}({\bf k}) = \delta_{ij}-{k_i k_j}/{k^2}$, $\varepsilon_{ijl}$
is the antisymmetric tensor, and $H^M\!(k)$ is the magnetic helicity spectrum.
{In this case, a white noise spectrum corresponds to the
spectral index $n=2$ \cite{hogan}, while the Batchelor spectrum corresponds
to the spectral index $n=4$ \cite{causal}. The power law of large-scale
MHD turbulence spectrum extends down to the integral scale $\xi_M$, which is
itself a time-dependent quantity throughout the turbulence decay process. At
scales shorter than the integral scale the spectral energy density of the
magnetic field decreases rapidly due to the combined action of turbulent
decay and viscous damping.}

\paragraph{Decay Laws.}

{The correlation length of the turbulent magnetic field
evolves in time during the free decay of turbulence. We may describe the
decay laws of the magnetic correlation length $\xi_M(\eta)$ and the
spectral energy density ${\mathcal E}_M(\eta)$ using two power law indices
$n_{\xi}$ and $n_E$:}
\begin{equation}
\xi_M(\eta) = \xi_M(\eta_0) \left({\eta \over \eta_0}\right)^{n_\xi} ~,
\label{xi(eta)}
\end{equation}
\begin{equation}
{\mathcal E}_M(\eta) = {\mathcal E}_M(\eta_0) \left({\eta \over \eta_0}\right)^{n_E}~.
\label{E(eta)}
\end{equation}
{The spectral energy density of the primordial MHD turbulence
spectrum can be split into its large-scale and short-scale components, above
and below the time dependent integral scale:}
\begin{equation}
E_M(k,\eta) = E_0(\eta)
\left\{ \begin{array}{l}
\bar k^4 ~~~~~~{\rm when}~~ k<k_I(\eta)  \\ \bar k^{-5/3} ~~{\rm when}~~ k>k_I(\eta)
\end{array} \right. ~,
\label{spectrum}
\end{equation}
where $\bar k = k / k_I$ and $k_I(\eta) = 2 \pi / \xi_M(\eta)$. Hence,
Eqs.~(6) and (7)
{can be used to describe the time evolution of spectral
amplitude of magnetic field $E_0(\eta)$ for a given turbulent spectrum:}
\begin{equation}
E_0(\eta) = {5 \over 17 \pi} \xi_M(\eta_0) {\mathcal E}_M(\eta_0)
\left({\eta \over \eta_0}\right)^{n_\xi + n_E} ~.
\label{E0(eta)}
\end{equation}

\section{Results}

\subsection{The Inverse Transfer for Non-Helical Fields}

\begin{figure}[t]
\includegraphics[width=0.9\columnwidth]{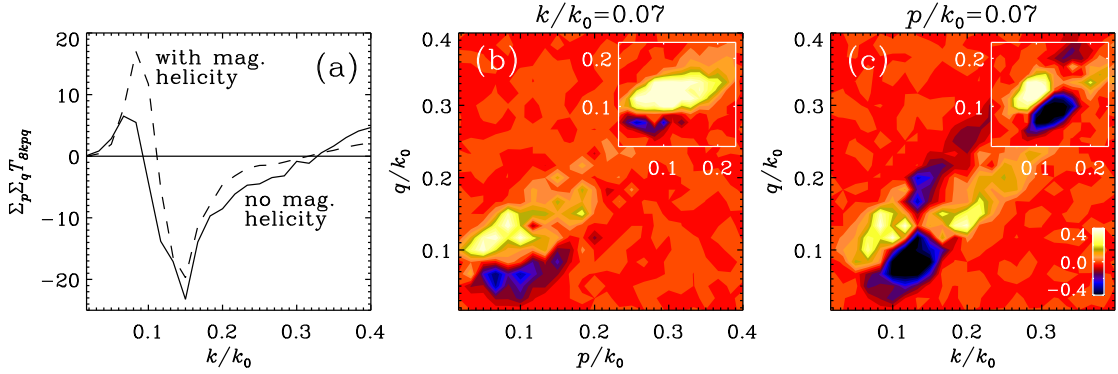}
\caption{Spectral transfer function $T_{kpq}$, (a) as a function of $k$ and
summed over all $p$ and $q$, (b) as a function of $p$ and $q$ for $k/k_1=4$,
and (c) as a function of $k$ and $q$ for $p/k_1=4$. The dashed line in (a)
and the insets in (b) and (c) show the corresponding case for a direct numerical
simulations with helicity; both for $\Pm=1$.
See Fig.~3 of Ref.~\cite{Brandenburg:2014mwa}.} \label{inverse-transfer}
\end{figure}

The inverse cascade is by now a well-known effect in helical magnetic
turbulence \cite{PFL}.
One of the remarkable results is the presence of {\em non-helical} inverse
transfer for magnetically dominated (causally generated) MHD turbulence;
see Fig.~\ref{inverse-transfer}, where we show spectral energy transfer rates,
which demonstrate that the inverse transfer is about half as strong as
with helicity. However, in both cases the magnetic gain at large scales results
from velocity at similar scales interacting with smaller-scale magnetic
fields \cite{Brandenburg:2014mwa}. This result has not been emphasized in
previous studies, see Refs.~\cite{Jedamzik:2010cy,Saveliev:2012ea} and
references therein, and has now been confirmed by independent research groups
\cite{Berera:2014hca,Zrake:2014mta,East:2015pea}.

Recent high resolution simulations with different magnetic Prandtl numbers
$\Pm=\nu/\lambda$ \cite{Brandenburg:2014mwa}
have shown a clear $k^{-2}$ spectrum in the
inertial range. This is the first example of fully isotropic magnetically
dominated MHD turbulence (governed by the phase transition-generated magnetic
fields) exhibiting what we argue to be weak turbulence scaling
\cite{Galtier:2000ce}. On the other hand, the Kolmogorov scaling $k^{-5/3}$
has been recovered for the case of the inflation-generated helical magnetic
fields \cite{bkt15}.

\subsection{Inflation Generated Magnetic Fields}

\begin{figure}[t]
\includegraphics[width=0.5 \columnwidth]{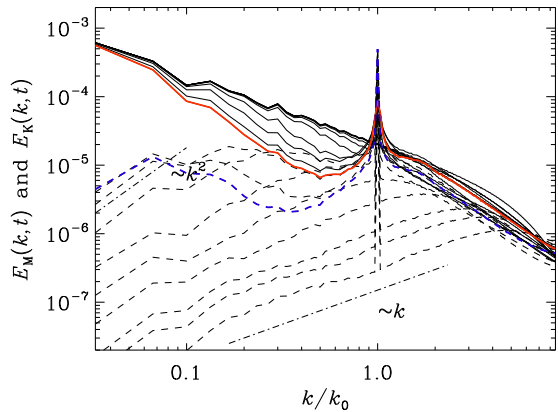}
\caption{
Magnetic (solid lines) and kinetic (dashed lines) energy spectra in
regular time intervals. $\Rey\approx170$.
The magnetic and kinetic spectra at the last time are additionally
marked in red and blue, respectively.
}\label{inflation}
\end{figure}

Magnetic fields generated during the inflation should be
affected by cosmological phase transitions occurring at later times during the
expansion of the universe. In this case, a separate study of the imprint of
phase transitions on cosmic magnetic fields is needed. For this purpose we
adopt a general approach that can be applied to both,
QCD and electroweak phase transitions. In each case, turbulence forcing is
determined by the phase transition bubble size. Rapid phase transitions
generate turbulence, which then decays slowly at large scales. In contrast to
previous studies, the inflation-generated magnetic field is not frozen into
the cosmic plasma. Turbulence is generated during a short forcing period,
which then is followed by slow decay (see
Refs.~\cite{Kahniashvili:2010gp,Kahniashvili:2012vt} for details). Recent
simulations showed an increasing characteristic length scale of the velocity
field and the establishment of a $k^2$ (white noise) spectrum at large
scales. This increase of vorticity perturbations occurs until it reaches
equipartition with the magnetic field \cite{Kahniashvili:2012uj}.
Figure~\ref{inflation} shows the evolution of kinetic and magnetic field
spectra from those simulations. Numerical results show that
inflation-generated magnetic fields are not significantly modified at large
scales by their coupling to the plasma during a cosmological phase
transition. The coupling of cosmic magnetic field with the phase
transition-generated fluid turbulence leads to deviations of the magnetic
field spectrum from the initial scale-invariant shape only at intermediate
scales.

Ongoing research consists in pursuing high resolution numerical
simulations of helical inflation-generated magnetic field evolution.
Such a field, being subject to inflationary expansion, is characterized by a
scale-invariant spectrum $n \rightarrow -1$, and its correlation length can be as
large as Hubble horizon today or even larger (i.e., even when the total
energy density ${\mathcal E}_M$ is finite, the correlation length $\xi_M
\propto \int dk E_M(k)/k$ divergences for $k \rightarrow 0$). In contrast to
well known helical magnetic field decay laws
\cite{b1,bII,b2,dav,Son:1998my,Saveliev:2013uva,campanelli,Christensson:2002xu,banerjee,
campanelli2,Christensson:2000sp}, {\it an absence} of the inverse cascade
has been found for inflation-generated magnetic fields.
Furthermore, an unusually slow growth of the correlation length and conservation
of helicity has been recovered even for the case of partially helical magnetic fields.
These unexpected and unknown features of magnetic helicity are the result of a
substantial turbulent power at large scales and the impossibility of the redistribution
of helical fields at small wavenumbers (only the forward cascade is possible).
A more thorough investigation of this phenomenon will be performed through
varying initial conditions and basic parameters of primordial plasma.

\begin{figure}[t]
\begin{center}
\includegraphics[width=0.9\textwidth]{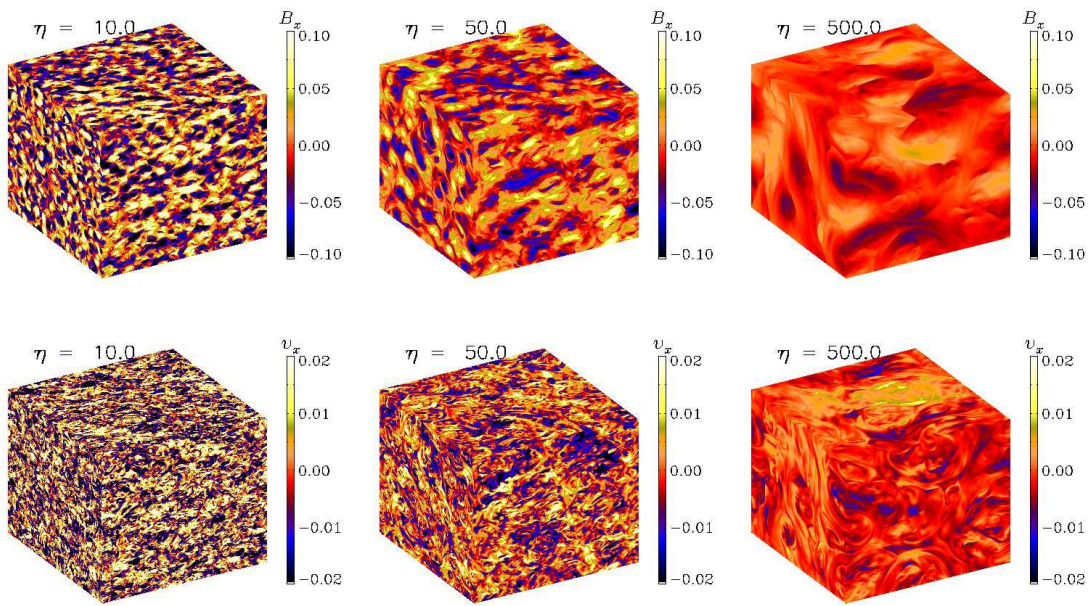}
%\plotone{f1}
\end{center}\caption[]{
Visualizations of $B_x$ (upper row) and $v_x$ (lower row) at three times
during the magnetic decay of a weakly helical field with
$\sigma=0.03$ generated during QCD phase transitions.
See Fig.~2 of Ref.~\cite{Tevzadze:2012kk}.} \label{BU}
\end{figure}

\begin{figure*}[t]%\begin{center}
\begin{minipage}[center]{0.8 \columnwidth}
\includegraphics[width=0.75\textwidth]{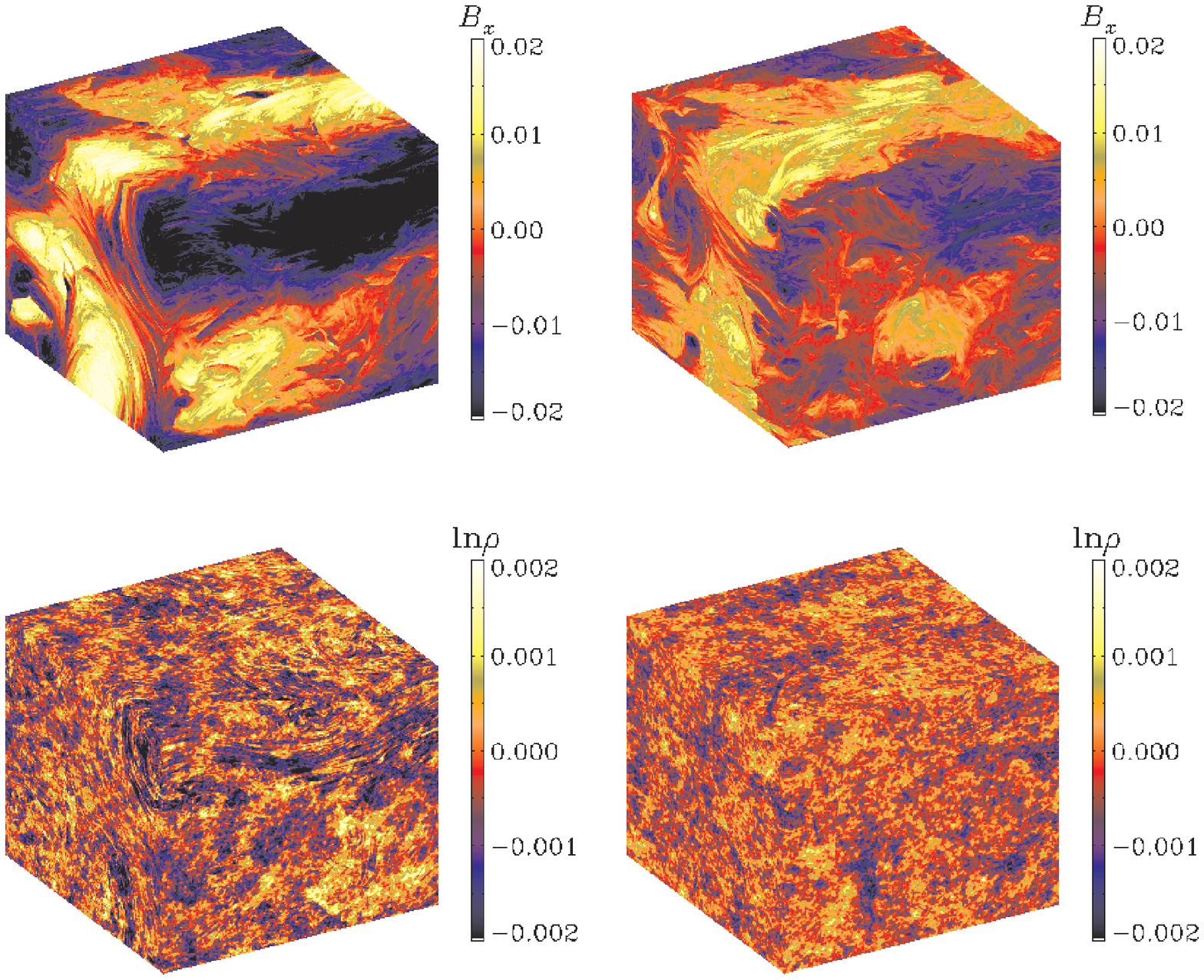}
\end{minipage}
\caption{ Comparison of $B_x$ (upper row) and $\ln\rho$ (lower row) for an
inflation-generated magnetic field with $\sigma=1$ (left) and $\sigma=0.03$
(right). }\label{bb1_sig003}
\end{figure*}

\subsection{Growth of Helical Structures}

{It is long known that the magnetic helicity plays a
crucial role in determining the evolution pattern of MHD turbulence. Distinct
evolution characteristics are known for helical and non-helical fields.
In recent simulations, a partially helical initial magnetic field was used
\cite{Tevzadze:2012kk}, assuming a tiny initial magnetic helicity during the
QCD phase transition.}
It was shown that at late times the resulting field attains the maximally
allowed magnetic helicity. This result is important since helicity crucially
affects the MHD dynamics, and has very interesting consequences in
astrophysical objects (e.g.\ galactic magnetic fields \cite{gal1,gal2,gal3},
for example). The resulting magnetic field has an amplitude of around
0.04\,nG and a correlation length of order 20\,kpc, which (assuming realistic
scenarios of amplification \cite{Dolag:2010ni}) serves as a seed for galactic
magnetic fields. At this point the electroweak phase transition-generated
magnetic fields are less promising due to a smaller initial correlation
length, but are not completely excluded \cite{Wagstaff:2014fla}, in
particular for fully helical magnetic fields \cite{Kahniashvili:2012uj}.

\begin{figure}[t!]
\includegraphics[width= 0.6 \textwidth]{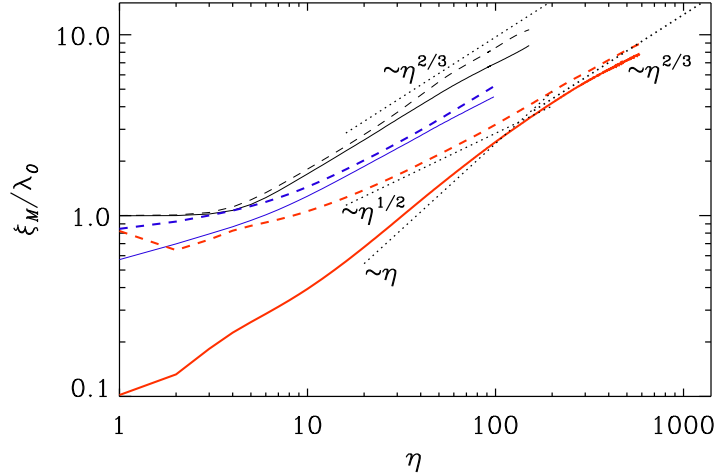}
\caption{
Evolution of turbulent correlation length $\xi_M(\eta)$
(solid) and minimal correlation length $\xi_M^{\rm min}(\eta)$ (dashed) of
a helical state for $\sigma$ = 1 (black), 0.1 (blue) and 0.03 (red).
(Fig.~3 of Ref.~\cite{Tevzadze:2012kk}).}
\label{hel-growth}
\end{figure}

{Magnetic helicity is a crucial factor that affects the
evolution of primordial magnetic fields. The evolution of the primordial
magnetic field that has been produced with weak initial magnetic helicity that
undergoes two consecutive stages. During the first stage, the evolution of a partially
helical magnetic field spectrum is much similar to that of
non-helical magnetic fields, and is sometimes described as a direct cascade.
At this stage the spectral energy density cascades from large to small
scales, where it undergoes de-correlation and viscous damping. Magnetic
helicity is conserved and hence its fractional value increases during the
turbulent decay process. The second stage in the primordial magnetic field
evolution sets in when the turbulent state with maximal helicity is reached.
The maximal value of the magnetic helicity that can be reached is limited by
the realizability condition. Indeed, the conservation of magnetic helicity
leads to a decay of the magnetic energy density inversely proportional to
the correlation length of the turbulence:}
\begin{equation}
\xi_M(\eta)\ge\xi_M^{\min}(\eta)\equiv|{\mathcal H}_M(\eta)|/2{\mathcal E}_M(\eta),
\end{equation}
{where $\xi_M^{\min}(\eta)$ is the minimal correlation length
of the turbulent state. Hence, an inverse cascade develops during the second
stage of the primordial magnetic field evolution.}
In Ref.~\cite{Tevzadze:2012kk}, we have studied the $\xi_M(\eta)$ and
$\xi_M^{\min}(\eta)$ for $\sigma=1$, $0.1$, and $0.03$ in the case of the QCD
phase transition.
{It seems that, especially at lower $\sigma$, the increase of
$\xi_M$ is slow ($\sim\eta^{1/2}$) while $\xi_M(\eta)\gg\xi_M^{\min}(\eta)$.}
{However, since magnetic helicity conservation implies that
${\mathcal E}_M$ decreases as $\eta^{-1}$, the minimal correlation length
$\xi_M^{\min}(\eta)$ soon reaches}
$\xi_M(\eta)$; see Fig.~(\ref{hel-growth}).
{When the correlation length of the turbulent magnetic field
reaches a minimal correlation length, the turbulence reaches its fully helical
state. Then the turbulence decays according to the helical turbulent decay
laws: $\xi_M\sim\eta^{2/3}$ and ${\mathcal E}_M\sim\eta^{-2/3}$.}
{Hence, we identify two distinct phases in the MHD turbulence
evolution: the phase of \emph{weakly helical} turbulence decay with $n_{\xi}
= 1/2$ and $n_E = -1$, and the phase of \emph{fully helical} turbulent decay
with  $n_{\xi} = 2/3$ and $n_E = -2/3$. The fully helical case is characterized
by an inverse cascade where $E_0(\eta) \propto \xi_M(\eta) {\mathcal E}_M(\eta) =
{\rm const}$ (see Eq.~(\ref{E0(eta)}). These results are in full agreement
with earlier works}
\cite{b1,b2, bII,dav,Son:1998my,Saveliev:2013uva,campanelli}.
The effective coupling of the primordial magnetic fields and
cosmic plasma ends at the time of recombination. At later stages, primordial magnetic
fields exhibit much slower developments \cite{beo96}.

{Knowing the initial values of the turbulent magnetic
correlation length $\xi_M(\eta_0)$, the minimal correlation length
$\xi_M^{\min}(\eta_0)$ set by the realizability condition, we can calculate the
time interval $\eta_{\rm fully}$ needed for the turbulence to reach its fully
helical state during the decay process.}
Since these two scales approach each other as $\eta^{1/2}$, the result is
$\eta_{\rm fully}=\eta_0[\xi_M(\eta_0)/\xi_M^{\min}(\eta_0)]^2$.
{Hence, the time interval needed for the development of a fully
helical state can be calculated using the initial values of turbulent
energy and helicity:}
\begin{equation}
\eta_{\rm fully}=4\eta_0 \xi_M^2 {\mathcal E}_M^2/{\mathcal H}_M^2.
\end{equation}
Note that this time increases
{inversely proportional to the square of the initial helicity of
the turbulent state ${\mathcal H}_M$.}
{Assuming that the initial magnetic helicity is $\xi_M /
\lambda_{H_\star}$  times lower then the maximal helicity in the case of strong
CP violation during phase transition, we can calculate the time needed for the
cosmic turbulence to reach its maximally helical state: $\eta_{\rm fully} =
\eta_0 / \gamma^2 $.}

\section{Discussion}

Let us now discuss our results in the broader context of the workshop
{\it `Mixing in Rapidly Changing Environments---Probing Matter at the
Extremes'}.
Our work has demonstrated a rather generic trend of decaying MHD
turbulence to display an increase of energy at large length scales.
This process is well-known in helical MHD turbulence \cite{PFL},
but to a lesser extent it also occurs in non-helical MHD turbulence
\cite{Brandenburg:2014mwa}.
Moreover, if there is small fractional helicity initially, this fraction
will increase with time proportional to the square root of time.
At a certain time, the fractional helicity will be 100\%, after which
the decay of energy slows down and the increase of the
correlation length speeds up.

An increase of energy at large length scales is not a common phenomenon
in hydrodynamic turbulence and has never been found for passive scalars.
This special behavior in MHD is likely to lead to unconventional mixing
properties, although this has not yet been well quantified for decaying
MHD turbulence.
However, for statistically stationary turbulence, an increase of
energy at small wavenumbers is usually described as non-diffusive
turbulent transport, which is particularly well known in mean-field
dynamo theory \cite{Blackman}.
In a sense, this is more reminiscent of what might look like ``anti-mixing''.
This is to some extent due to the fact that vector fields behave
differently from scalar fields.
This can in part be due to the presence of additional conservation laws.
In particular, magnetic helicity provides an extremely powerful constraint.

The significance of studying decaying MHD turbulence is manifold.
On the one hand, our results will help to better understand the nature
of cosmic magnetism and will gain insight into the decay laws of cosmic
MHD turbulence.
On the other hand, our analysis can be
applied not only to cosmological scales, but to molecular clouds or even
protoplanetary disks where decaying magnetic turbulence can crucially affect
the global state or the formation of local structures.
The results of our research have therefore important implications in many areas,
including fluid dynamics, early universe physics, high energy astrophysics,
MHD modeling, and large-scale structure formation in the universe.

\section{Conclusions}

We have discussed the evolution of primordial magnetic fields during the
expansion of the universe and have addressed some of the observational
signatures.
The coupling between the magnetic fields and cosmic turbulence
leads to novel results as compared to previously adopted frozen-in
approximations when magnetic field evolution was considered solely due to the
field line dilution in the expanding universe. For this purpose, we have
developed a comprehensive theoretical model to consider the evolution of MHD
turbulence over a wide range of physical conditions, beyond any specific
astrophysical application.

Numerical simulations have shown novel effects in the evolution of
magnetic fields in cosmic turbulence. It seems that inverse transfer that
normally occurs in helical MHD turbulence, can also take place for
non-helical magnetic fields if the MHD turbulence is magnetically dominated.
In this case, large-scale velocity perturbations power up the magnetic field,
leading to substantial increase of magnetic power spectra at large scales and
corresponding inverse transfer.

An analysis of the inflation-generated magnetic field is carried out to find out
how they are affected by the cosmic phase transitions (QCD and electroweak
phase transition). Numerical results show that large-scale magnetic fields
survive phase transitions, and thus, phase transitions cannot rule out
inflationary magnetogenesis as the source of seed magnetic field in the
universe.

On the other hand, magnetic fields generated during phase transitions can have
tiny helicity.
We have shown that during the evolution in MHD turbulence magnetic field helicity
grows until it reaches maximal helical state. It seems that helicity growth
rate is fast enough to reach maximal helicity well before the epoch of
recombination, when the primordial magnetic field decouples from cosmic turbulence.

We have used results of high resolution simulations to set limits on the
large-scale magnetic field generated during early stages of the evolution
of the universe (inflation or cosmological phase transitions). Indeed, it seems
that magnetic fields produced during this epochs and subsequently modified by
cosmic MHD turbulence
can reach amplitudes similar to the lower bounds of the
observational magnetic fields even accounting for the effects of large-scale
turbulent decay as well as additional Alfv\'en wave damping.
The extremely low values
{derived for smoothed magnetic field \cite{Caprini:2001nb} do
not imply that the effective magnetic field}
{is also small in the $1\,$pc--$1\,$kpc range and cannot lead
to the observational signatures in blazar emission spectra.}
Using the effective magnetic field approach we obtain results that do not
depend on the specific spectral shape of the magnetized turbulence.
{Observational signatures of the magnetogenesis during
electroweak phase transition can come from future observations, if weak
magnetic fields with $10^{-14}$--$10^{-15}$\,G amplitude and few pc
correlation length are detected.}
While a somewhat stronger magnetic field with a correlation length of the
order of kpc might indicate the presence of QCD phase transition
magnetogenesis.
{In turn, magnetic fields} with extremely large correlation
length (1Mpc or higher) will indicate inflationary magnetogenesis.

\begin{figure}[th]
\includegraphics[width= 0.6 \textwidth]{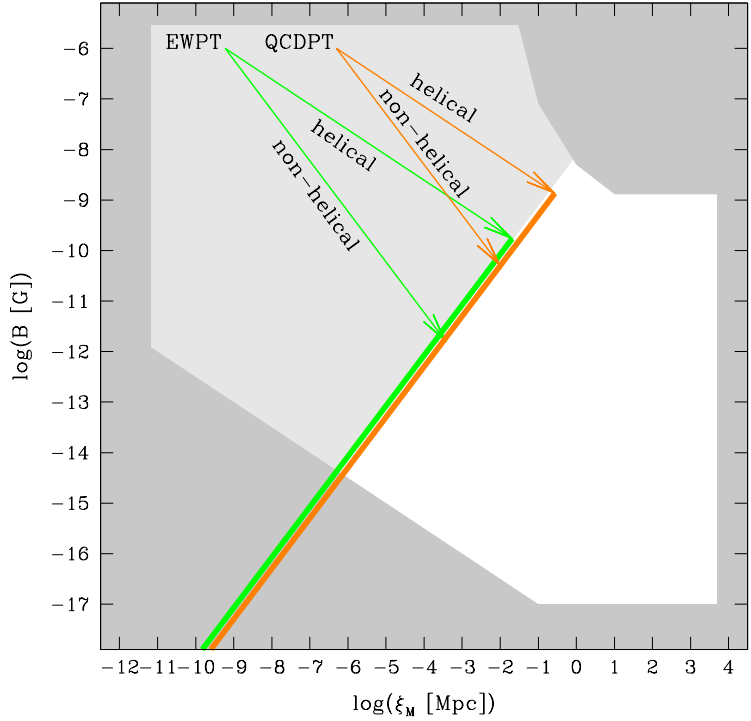}
\caption{
Evolution of the effective magnetic field $B^{\rm (eff)}$
and correlation length $\xi_M$ in the case of magnetogenesis at electroweak
phase transition (green) and QCD phase transition (orange).
Arrows indicate the evolutionary path of the strength
and integral scale of helical and non-helical turbulent magnetic fields
during the radiation-dominated era up to their final values.
Thick solid line(s) show possible present day field strengths and
integral scales of the magnetic field generated during phase transitions
(see Fig.~7 of Ref.~\cite{Kahniashvili:2012uj}).}
\label{fig:nonhelical}
\end{figure}

\acknowledgments T.K. would like to express her appreciation to Organizing
Committee of Turbulent Mixing and Beyond Workshop {\it
'Mixing in Rapidly Changing Environments - Probing Matter at the Extremes'}
(2014, The Abdus Salam  International Centre for Theoretical Physics, ICTP) and the Center for
hospitality.  It is our pleasure to thank Marco Ajello, Nick Battaglia, Leonardo
Campanelli, Rupert Croft, Ruth Durrer,  Arthur Kosowsky, Francesco Miniati, Aravind Natarjan, Andrii Neronov,
Yurii Maravin, and Tanmay Vachaspati for useful discussions. We acknowledge
partial support from the Swedish Research Council grants 621-2011-5076 and
2012-5797, the European Research Council AstroDyn Research Project 227952,
the FRINATEK grant 231444 under the Research Council of Norway, the Swiss NSF
SCOPES grant IZ7370-152581, the NASA Astrophysics Theory  program grant
NNXl0AC85G, the NSF Astrophysics and Astronomy Grant Program grants AST-1109180,
AST-1615100 and AST-1615940,
and Shota Rustaveli Georgian National Science Foundation grant FR/264/6-350/14.


\begin{thebibliography}{widest-label}
\expandafter\ifx\csname natexlab\endcsname\relax\def\natexlab#1{#1}\fi
\expandafter\ifx\csname natexlab\endcsname\relax\def\natexlab#1{#1}\fi

\bibitem{Widrow:2002ud}
  L.~M.~Widrow,
%  ``Origin of galactic and extragalactic magnetic fields,''
  Rev.\ Mod.\ Phys.\  {\bf 74}, 775 (2002).
  %[astro-ph/0207240].
  %%CITATION = ASTRO-PH/0207240;%%
  %336 citations counted in INSPIRE as of 08 Nov 2014

\bibitem{Vallee} J.\ P.\ Vall{\'e}e,
%"Cosmic magnetic fields---as observed in the Universe, in galactic dynamos, and in the Milky Way",
New Astron.\ Rev.\ {\bf 48}, 763 (2004).

\bibitem{Giovannini:2006kg}
  M.~Giovannini,
 % ``Magnetic fields, strings and cosmology,''
  Lect.\ Notes Phys.\  {\bf 737}, 863 (2008).
  %[astro-ph/0612378].
  %%CITATION = ASTRO-PH/0612378;%%
  %33 citations counted in INSPIRE as of 08 Nov 2014

\bibitem{Beck:2011gv} R.\ Beck, AIP Conf.\ Proc.\  {\bf 1085}, 83 (2009);
    R.~Beck,
  %``Future Observations of Cosmic Magnetic Fields with the SKA and its Precursors,''
  In Proc. of Magnetic Fields in the Universe: From Laboratory and Stars to Primordial
  Structures, eds. M. Soida, K. Otmianowska-Mazur,
  E.M. de Gouveia Dal Pina and A. Lazarian; SKA design and timeline updates, arXiv:1111.5802
  [astro-ph.CO].

\bibitem{VanEck:2010ka}
  C.~Van Eck,
  %J.~-A.~Brown, J.~Stil, K.~Rae, S.~A.~Mao, B.~Gaensler, A.~Shukurov, R.~Taylor, M. Haverkorn, P. Kronberg, N. McClure-Griffiths,
  {\it et al.},
%``Modeling the Magnetic Field in the Galactic Disk using New Rotation Measure Observations from the Very Large Array,''
  Astrophys.\ J.\  {\bf 728}, 97 (2011).
    %[arXiv:1012.2938 [astro-ph.GA]].  %%CITATION = ARXIV:1012.2938;%%

%\cite{Fletcher:2010wt}
\bibitem{Fletcher:2010wt}
  A.~Fletcher, R.~Beck, A.~Shukurov, E.~M.~Berkhuijsen and C.~Horellou,
%  ``Magnetic fields and spiral arms in the galaxy M51,'',
Mon.\ Not.\ Roy.\ Astron.\ Soc.\ {\bf 112}, 2396 (2011).



\bibitem{Beck:2011he} R.~Beck,
% ``Cosmic Magnetic Fields: Observations and Prospects,''
  AIP Conf.\ Proc.\  {\bf 1381}, 117 (2011). %  [ [astro-ph.CO]].

\bibitem{Bernet:2008qp}
  M.~L.~Bernet, F.~Miniati, S.~J.~Lilly, P.~P.~Kronberg and M.~Dessauges-Zavadsky,
%  ``Strong magnetic fields in normal galaxies at high redshifts,''
  Nature {\bf 454}, 302 (2008). %[arXiv:0807.3347 [astro-ph]].
  %%CITATION = ARXIV:0807.3347;%%
  %128 citations counted in INSPIRE as of 06 Nov 2015

\bibitem{Kronberg:2007dy}
  P.~P.~Kronberg, M.~L.~Bernet, F.~Miniati, S.~J.~Lilly, M.~B.~Short and D.~M.~Higdon,
%  ``A Global Probe of Cosmic Magnetic Fields to High Redshifts,''
  Astrophys.\ J.\  {\bf 676}, 7079 (2008).%[arXiv:0712.0435 [astro-ph]].
  %%CITATION = ARXIV:0712.0435;%%
  %26 citations counted in INSPIRE as of 06 Nov 2015

\bibitem{Durrer:2013pga}
  R.~Durrer and A.~Neronov,
%  ``Cosmological Magnetic Fields: Their Generation, Evolution and Observation,''
  Astron.\ Astrophys.\ Rev.\  {\bf 21}, 62 (2013).
  %[arXiv:1303.7121 [astro-ph.CO]].
  %%CITATION = ARXIV:1303.7121;%%
  %66 citations counted in INSPIRE as of 09 Nov 2014

\bibitem{Kulsrud:2007an}
  R.~M.~Kulsrud and E.~G.~Zweibel,
%  ``The Origin of Astrophysical Magnetic Fields,''
  Rept.\ Prog.\ Phys.\  {\bf 71}, 0046091 (2008).



\bibitem{Kandus:2010nw}
  A.~Kandus, K.~E.~Kunze and C.~G.~Tsagas,
%  ``Primordial magnetogenesis,''
  Phys.\ Rept.\  {\bf 505}, 1 (2011).

\bibitem{b1} D. Biskamp, {\it  Magnetohydrodynamic Turbulence } (Cambridge: Cambridge
Univ. Press) 2003.

\bibitem{Kravtsov:2012zs}
A.~Kravtsov and S.~Borgani,
%``Formation of Galaxy Clusters,''
  Ann. Rev.  Astron. Astrophys., {\bf 50}, 353 (2012).
  %arXiv:1205.5556 [astro-ph.CO].

%\cite{Subramanian:2005hf}
\bibitem{Subramanian:2005hf}
K.~Subramanian, A.~Shukurov and N.~E.~L.~Haugen,
%  ``Evolving turbulence and magnetic fields in galaxy clusters,''
  Mon.\ Not.\ Roy.\ Astron.\ Soc.\  {\bf 366}, 1437 (2006).

%\cite{Vazza:2006cp}
\bibitem{Vazza:2006cp}
  F.~Vazza, G.~Tormen, R.~Cassano, G.~Brunetti and K.~Dolag,
%  ``Turbulent velocity fields in sph-simulated galaxy clusters,''
  Mon.\ Not.\ Roy.\ Astron.\ Soc.\  {\bf 369}, L14 (2006).%[astro-ph/0602247].
  %%CITATION = ASTRO-PH/0602247;%%
  %48 citations counted in INSPIRE as of 11 Nov 2015

%\cite{Fang:2008ad}
\bibitem{Fang:2008ad}
  T.~Fang, P.~J.~Humphrey and D.~A.~Buote,
%  ``Rotation and Turbulence of the Hot ICM in Galaxy Clusters,''
  Astrophys.\ J.\  {\bf 691}, 1648 (2009). %[arXiv:0808.1106 [astro-ph]].
  %%CITATION = ARXIV:0808.1106;%%
  %42 citations counted in INSPIRE as of 11 Nov 2015


\bibitem{Greif:2008qqa}
T.~H.~Greif, J.~L.~Johnson, R.~S.~Klessen and V.~Bromm,
%  ``The First Galaxies: Assembly, Cooling and the Onset of Turbulence,''  %arXiv:0803.2237 [astro-ph];
  Mon. Not. Roy. Astron. Soc. {\bf 387}, 1021 (2008).

\bibitem{Pakmor:2011ht}
  R.~Pakmor, A.~Bauer and V.~Springel,
%  ``Magnetohydrodynamics on an unstructured moving grid,''
  Mon. Not. Roy. Astron. Soc. {\bf  418}, 1392 (2012).
  %arXiv:1108.1792 [astro-ph.IM];


%\cite{Vazza:2014jga}
\bibitem{Vazza:2014jga}
  F.~Vazza, M.~Br�ggen, C.~Gheller and P.~Wang,
%  ``On the amplification of magnetic fields in cosmic filaments and galaxy clusters,''
  Mon.\ Not.\ Roy.\ Astron.\ Soc.\  {\bf 445}, no. 4, 3706 (2014).%  [arXiv:1409.2640 [astro-ph.CO]].
  %%CITATION = ARXIV:1409.2640;%%
  %8 citations counted in INSPIRE as of 11 Nov 2015

\bibitem{Cho:2014oca}
  J.~Cho,
%  ``Origin of Magnetic Field in the Intracluster Medium: Primordial or Astrophysical?,''
  Astrophys.\ J.\  {\bf 797}, no. 2, 133 (2014). %[arXiv:1410.1893 [astro-ph.CO]].
  %%CITATION = ARXIV:1410.1893;%%
  %6 citations counted in INSPIRE as of 11 Nov 2015

\bibitem{Espinosa:2007}
    J.~R.~Espinosa, T.~Konstandin, J.~M.~No and G.~Servant,
%     `The spectrum of gravitational radiation from primordial turbulence,''
  Phys.\ Rev.\  D {\bf 76}, 083002 (2007).

%\cite{Espinosa:2010hh}
\bibitem{Espinosa:2010hh}
  J.~R.~Espinosa, T.~Konstandin, J.~M.~No and G.~Servant,
%  ``Energy Budget of Cosmological First-order Phase Transitions,''
  JCAP {\bf 1006}, 028 (2010).
    % [arXiv:1004.4187 [hep-ph]].  %%CITATION = ARXIV:1004.4187;%%



\bibitem{Medina:2014}
S.-N. X. Medina, S. J. Arthur, W. J. Henney, G. Mellema, A. Gazol,
%"Turbulence in simulated H II regions"
Mon. Not. Roy. Astron. Soc. {\bf 445}, 1797 (2014).
%1409.3858

\bibitem{Neronov:1900zz}
  A.~Neronov and I.~Vovk,
%  ``Evidence for strong extragalactic magnetic fields from Fermi observations of TeV blazars,''
  Science {\bf 328}, 73 (2010). %[arXiv:1006.3504 [astro-ph.HE]].
  %%CITATION = ARXIV:1006.3504;%%
  %312 citations counted in INSPIRE as of 06 Nov 2015

\bibitem{Tavecchio:2010ja}
  F.~Tavecchio, G.~Ghisellini, G.~Bonnoli and L.~Foschini,
%  ``Extreme TeV blazars and the intergalactic magnetic field,''
  Mon.\ Not.\ Roy.\ Astron.\ Soc.\  {\bf 414}, 3566 (2011). %[arXiv:1009.1048 [astro-ph.HE]].
  %%CITATION = ARXIV:1009.1048;%%
  %64 citations counted in INSPIRE as of 06 Nov 2015


%\cite{Essey:2010nd}
\bibitem{Essey:2010nd}
  W.~Essey, S.~Ando and A.~Kusenko,
%  ``Determination of intergalactic magnetic fields from gamma ray data,''
  Astropart.\ Phys.\  {\bf 35}, 135 (2011).%[arXiv:1012.5313 [astro-ph.HE]].
  %%CITATION = ARXIV:1012.5313;%%
  %74 citations counted in INSPIRE as of 06 Nov 2015

%\cite{Taylor:2011bn}
\bibitem{Taylor:2011bn}
  A.~M.~Taylor, I.~Vovk and A.~Neronov,
%  ``Extragalactic magnetic fields constraints from simultaneous GeV-TeV observations of blazars,''
  Astron.\ Astrophys.\  {\bf 529}, A144 (2011). %[arXiv:1101.0932 [astro-ph.HE]].
  %%CITATION = ARXIV:1101.0932;%%
  %91 citations counted in INSPIRE as of 06 Nov 2015


%\cite{Huan:2011kp}
\bibitem{Huan:2011kp}
  H.~Huan, T.~Weisgarber, T.~Arlen and S.~P.~Wakely,
%  ``A New Model for Gamma-Ray Cascades in Extragalactic Magnetic Fields,''
  Astrophys.\ J.\  {\bf 735}, L28 (2011). %[arXiv:1106.1218 [astro-ph.HE]].
  %%CITATION = ARXIV:1106.1218;%%
  %14 citations counted in INSPIRE as of 11 Nov 2015

%\cite{Vovk:2011aa}
\bibitem{Vovk:2011aa}
  I.~Vovk, A.~M.~Taylor, D.~Semikoz and A.~Neronov,
%  ``Fermi/LAT observations of 1ES 0229+200: implications for extragalactic magnetic fields and background light,''
  Astrophys.\ J.\  {\bf 747}, L14 (2012). %[arXiv:1112.2534 [astro-ph.CO]].
  %%CITATION = ARXIV:1112.2534;%%
  %49 citations counted in INSPIRE as of 06 Nov 2015

%\cite{Dolag:2010ni}
\bibitem{Dolag:2010ni}
  K.~Dolag, M.~Kachelriess, S.~Ostapchenko and R.~Tomas,
%  ``Lower limit on the strength and filling factor of extragalactic magnetic fields,''
  Astrophys.\ J.\  {\bf 727}, L4 (2011).%[arXiv:1009.1782 [astro-ph.HE]].
  %%CITATION = ARXIV:1009.1782;%%
  %122 citations counted in INSPIRE as of 11 Nov 2015


\bibitem{Dermer:2010mm}
  C.~D.~Dermer, M.~Cavadini, S.~Razzaque, J.~D.~Finke and B.~Lott,
%  ``Time Delay of Cascade Radiation for TeV Blazars and the Measurement of the Intergalactic Magnetic Field,''
  Astrophys.\ J.\  {\bf 733}, L21 (2011). %[arXiv:1011.6660 [astro-ph.HE]].
  %%CITATION = ARXIV:1011.6660;%%
  %97 citations counted in INSPIRE as of 11 Nov 2015



%\cite{Takahashi:2011ac}
\bibitem{Takahashi:2011ac}
  K.~Takahashi, M.~Mori, K.~Ichiki and S.~Inoue,
%  ``Lower Bounds on Intergalactic Magnetic Fields from Simultaneously Observed GeV-TeV Light Curves of the Blazar Mrk 501,''
  Astrophys.\ J.\  {\bf 744}, L7 (2012).%[arXiv:1103.3835 [astro-ph.CO]].
  %%CITATION = ARXIV:1103.3835;%%
  %35 citations counted in INSPIRE as of 11 Nov 2015


\bibitem{Finke:2015ona}
  J.~D.~Finke, L.~C.~Reyes, M.~Georganopoulos, K.~Reynolds, M.~Ajello, S.~J.~Fegan and K.~McCann,
%  ``Constraints on the Intergalactic Magnetic Field with Gamma-Ray Observations of Blazars,''
  Astrophys.\ J.\  {\bf 814}, no. 1, 20 (2015).
  %arXiv:1510.02485 [astro-ph.HE].
  %%CITATION = ARXIV:1510.02485;%%

\bibitem{Arlen:2012iy}
  T.~C.~Arlen, V.~V.~Vassiliev, T.~Weisgarber, S.~P.~Wakely and S.~Y.~Shafi,
%  ``Intergalactic Magnetic Fields and Gamma Ray Observations of Extreme TeV Blazars,''
  Astrophys.\ J.\  {\bf 796}, 18 (2014).
  %arXiv:1210.2802 [astro-ph.HE].
%  %%CITATION = ARXIV:1210.2802;%%

\bibitem{Broderick:2011ab}
  A.\ E.\ Broderick, P.\ Chang and C.\ Pfrommer,
%  ``The Cosmological Impact of Luminous TeV Blazars I: Implications of Plasma Instabilities for the Intergalactic Magnetic Field and Extragalactic Gamma-Ray Background,''
  Astrophys.\ J.\  {\bf 752}, 22 (2012).

%\cite{Miniati:2012ge}
\bibitem{Miniati:2012ge}
  F.~Miniati and A.~Elyiv,
%``Relaxation of Blazar Induced Pair Beams in Cosmic Voids: Measurement of Magnetic Field in Voids and Thermal History of the IGM,''
  Astrophys.\ J.\  {\bf 770}, 54 (2013).
  %[arXiv:1208.1761 [astro-ph.CO]].
  %%CITATION = ARXIV:1208.1761;%%
  %29 citations counted in INSPIRE as of 09 Nov 2014


\bibitem{Lemoine:2014gca}
  M.~Lemoine,
%  ``Non-linear collisionless damping of Weibel turbulence in relativistic blast waves,''
 J.\ Plasma Phys.\  {\bf 81}, 4501 (2015).
%  arXiv:1410.0146 [physics.plasm-ph].
  %%CITATION = ARXIV:1410.0146;%%

%%%%%%%%%%%%%%%%%%%%%%%%%%%%%%%%%%%%%%%%%%%%%%%%%%%%%%%%%%%%%%%%%%%%%%%%%%%%%%%%%

%\cite{Kosowsky:1996yc}
\bibitem{Kosowsky:1996yc}
  A.~Kosowsky and A.~Loeb,
%  ``Faraday rotation of microwave background polarization by a primordial magnetic field,''
  Astrophys.\ J.\  {\bf 469}, 1 (1996). %[astro-ph/9601055].
  %%CITATION = ASTRO-PH/9601055;%%
  %191 citations counted in INSPIRE as of 11 Nov 2015

%\cite{Harari:1996ac}
\bibitem{Harari:1996ac}
  D.~D.~Harari, J.~D.~Hayward and M.~Zaldarriaga,
%  ``Depolarization of the cosmic microwave background by a primordial magnetic field and its effect upon temperature anisotropy,''
  Phys.\ Rev.\ D {\bf 55}, 1841 (1997). %[astro-ph/9608098].
  %%CITATION = ASTRO-PH/9608098;%%
  %54 citations counted in INSPIRE as of 11 Nov 2015

%\cite{Scoccola:2004ke}
\bibitem{Scoccola:2004ke}
  C.~Scoccola, D.~Harari and S.~Mollerach,
%  ``B polarization of the CMB from Faraday rotation,''
  Phys.\ Rev.\ D {\bf 70}, 063003 (2004)
  [astro-ph/0405396].
  %%CITATION = ASTRO-PH/0405396;%%
  %49 citations counted in INSPIRE as of 11 Nov 2015

%\cite{Campanelli:2004pm}
\bibitem{Campanelli:2004pm}
  L.~Campanelli, A.~D.~Dolgov, M.~Giannotti and F.~L.~Villante,
%  ``Faraday rotation of the CMB polarization and primordial magnetic field properties,''
  Astrophys.\ J.\  {\bf 616}, 1 (2004).%[astro-ph/0405420].
  %%CITATION = ASTRO-PH/0405420;%%
  %81 citations counted in INSPIRE as of 11 Nov 2015

\bibitem{Kosowsky:2004zh} A.~Kosowsky, T.~Kahniashvili, G.~Lavrelashvili
 and B.~Ratra,
% ``Faraday Rotation of the Cosmic Microwave Background Polarization by a Stochastic Magnetic Field,''
  Phys.\ Rev.\  D {\bf 71}, 043006 (2005).


\bibitem{Kahniashvili:2008hx}  T.~Kahniashvili, Y.~Maravin and A.~Kosowsky,
%  ``Faraday Rotation Limits On A Primordial Magnetic Field From Wilkinson Microwave Anisotropy Probe Five-Year Data,''
  Phys.\ Rev.\  D {\bf 80}, 023009 (2009).

\bibitem{Giovannini:2008aa}
  M.~Giovannini and K.~E.~Kunze,
%  ``Faraday rotation, stochastic magnetic fields and CMB maps,''
  Phys.\ Rev.\ D {\bf 78}, 023010 (2008).%[arXiv:0804.3380 [astro-ph]].
  %%CITATION = ARXIV:0804.3380;%%
  %50 citations counted in INSPIRE as of 11 Nov 2015


\bibitem{Kahniashvili:2010wm} T.~Kahniashvili, A.~G.~Tevzadze, S.~K.~Sethi, K.~Pandey and B.~Ratra,
%  ``Primordial magnetic field limits from cosmological data,''
  Phys.\ Rev.\  D {\bf 82}, 083005 (2010).
%arXiv:1009.2094 [astro-ph.CO].

\bibitem{Pogosian:2011qv}
  L.~Pogosian, A.~P.~S.~Yadav, Y.~F.~Ng and T.~Vachaspati,
%  ``Primordial Magnetism in the CMB: Exact Treatment of Faraday Rotation and WMAP7 Bounds,''
  Phys.\ Rev.\ D {\bf 84}, 043530 (2011).
  %[Erratum-ibid.\ D {\bf 84}, 089903 (2011)]
  %[arXiv:1106.1438 [astro-ph.CO]].
  %%CITATION = ARXIV:1106.1438;%%
  %20 citations counted in INSPIRE as of 08 Nov 2014



%\cite{Ade:2015cao}
\bibitem{Ade:2015cao}
  P.~A.~R.~Ade,
  %K. Arnold, M. Atlas, C. Baccigalupi, D. Barron, D. Boettger, J. Borrill, S. Chapman, Y. Chinone, A. Cukierman, M. Dobbs, A. Ducout, R. Dunner, T. Elleflot, J. Errard, G. Fabbian, S. Feeney, C. Feng, A. Gilbert, N. Goeckner-Wald, J. Groh, G. Hall, N. W. Halverson, M. Hasegawa, K. Hattori, M. Hazumi, C. Hill, W. L. Holzapfel, Y. Hori, L. Howe, Y. Inoue, G. C. Jaehnig, A. H. Jaffe, O. Jeong, N. Katayama, J. P. Kaufman, B. Keating, Z. Kermish, R. Keskitalo, T. Kisner, A. Kusaka, M. Le Jeune, A. T. Lee, E. M. Leitch, D. Leon, Y. Li, E. Linder, L. Lowry, F. Matsuda, T. Matsumura, N. Miller, J. Montgomery, M. J. Myers, M. Navaroli, H. Nishino, T. Okamura, H. Paar, J. Peloton, L. Pogosian, D. Poletti, G. Puglisi, C. Raum, G. Rebeiz, C. L. Reichardt, P. L. Richards, C. Ross, K. M. Rotermund, D. E. Schenck, B. D. Sherwin, M. Shimon, I. Shirley, P. Siritanasak, G. Smecher, N. Stebor, B. Steinbach, A. Suzuki, J. Suzuki, O. Tajima, S. Takakura, A. Tikhomirov, T.  Tomaru, N. Whitehorn, B. Wilson, A. Yadav, A. Zahn, O. Zahn,
   {\it et al.}
  [POLARBEAR Collaboration],
%  ``POLARBEAR Constraints on Cosmic Birefringence and Primordial Magnetic Fields,''
  arXiv:1509.02461 [astro-ph.CO].
  %%CITATION = ARXIV:1509.02461;%%
  %2 citations counted in INSPIRE as of 11 Nov 2015

%\cite{Pshirkov:2015tua}
\bibitem{Pshirkov:2015tua}
  M.~S.~Pshirkov, P.~G.~Tinyakov and F.~R.~Urban,
 % ``New limits on extragalactic magnetic fields from rotation measures,''
  arXiv:1504.06546 [astro-ph.CO].
  %%CITATION = ARXIV:1504.06546;%%
  %2 citations counted in INSPIRE as of 11 Nov 2015


%%%%%%%%%%%%%%%%%%%%%%%%%%%%%%%%%%%%%%%%%%%%%%%%%%%%%%%%%%%%%%%%%%%%%%%%%%%%%%%%%%%%%%%%%%%%%%%%%%%%%%

\bibitem{a1} P. P. Kronberg, M. Simard-Normandin,
%"New evidence on the origin of rotation measures in extragalactic radio sources",
Nature, {\bf 263}, 653
(1976).

\bibitem{a11}
 P. P.  Kronberg, J. J. Perry,
%"Absorption lines, Faraday rotation, and magnetic field estimates for QSO absorption-line clouds",
Astrophys. J. {\bf 263}, 518 (1982).

\bibitem{a2} P. Blasi,  S.  Burles, A. V. Olinto
%"Cosmological Magnetic Field Limits in an Inhomogeneous Universe",
Astrophys. J. {\bf 514}, L79, (1999).

\bibitem{a3}
E.~Komatsu,
%J. Dunkley, M. R. Nolta, C. L. Bennett, B. Gold, G. Hinshaw, N. Jarosik, D. Larson, M. Limon, L. Page, D. N. Spergel, M. Halpern, R. S. Hill, A. Kogut, S. S. Meyer, G. S. Tucker, J. L. Weiland, E. Wollack, E. L. Wright,
  %
  {\it et al.}
  [WMAP Collaboration],
%``Five-Year Wilkinson Microwave Anisotropy Probe (WMAP) Observations: Cosmological Interpretation,''
  Astrophys.\ J.\ Suppl.\  {\bf 180}, 330 (2009).



%%%%%%%%%%%%%%%%%%%%%%%%%%%%%%%%%%%%%%%%%%%%%%%%%%%%%%%%%%%%%%%%%%%%%%%%%%%%%%%%%%
\bibitem{Ade:2015cva}
  {\it et al.}
%P. A. R. Ade, N. Aghanim, M. Arnaud, F. Arroja, M. Ashdown, J. Aumont, C. Baccigalupi, M. Ballardini, A. J. Banday, R. B. Barreiro, N. Bartolo, E. Battaner, K. Benabed, A. Beno�t, A. Benoit-L�vy, J.-P. Bernard, M. Bersanelli, P. Bielewicz, A. Bonaldi, L. Bonavera, J. R. Bond, J. Borrill, F. R. Bouchet, M. Bucher, C. Burigana, R. C. Butler, E. Calabrese, J.-F. Cardoso, A. Catalano, A. Chamballu, H. C. Chiang, J. Chluba, P. R. Christensen, S. Church, D. L. Clements, S. Colombi, L. P. L. Colombo, C. Combet, F. Couchot, A. Coulais, B. P. Crill, A. Curto, F. Cuttaia, L. Danese, R. D. Davies, R. J. Davis, P. de Bernardis, A. de Rosa, G. de Zotti, J. Delabrouille, F.-X. D�sert, J. M. Diego, K. Dolag, H. Dole, S. Donzelli, O. Dor�, M. Douspis, A. Ducout, X. Dupac, G. Efstathiou, F. Elsner, T. A. En�lin, H. K. Eriksen, J. Fergusson, F. Finelli, E. Florido, O. Forni, M. Frailis, A. A. Fraisse, E. Franceschi, A. Frejsel, S. Galeotta, S. Galli, K. Ganga, M. Giard, Y. Giraud-H�raud, E. Gjerl�w, J. Gonz�lez-Nuevo, K. M. G�rski, S. Gratton, A. Gregorio, A. Gruppuso, J. E. Gudmundsson, F. K. Hansen, D. Hanson, D. L. Harrison, G. Helou, S. Henrot-Versill�, C. Hern�ndez-Monteagudo, D. Herranz, S. R. Hildebrandt, E. Hivon, M. Hobson, W. A. Holmes, A. Hornstrup, W. Hovest, K. M. Huffenberger, G. Hurier, A. H. Jaffe, T. R. Jaffe, W. C. Jones, M. Juvela, E. Keih�nen, R. Keskitalo, J. Kim, T. S. Kisner, J. Knoche, M. Kunz, H. Kurki-Suonio, G. Lagache, A. L�hteenm�ki, J.-M. Lamarre, A. Lasenby, M. Lattanzi, C. R. Lawrence, J. P. Leahy, R. Leonardi, J. Lesgourgues, F. Levrier, M. Liguori, P. B. Lilje, M. Linden-V�rnle, M. L�pez-Caniego, P. M. Lubin, J. F. Mac�as-P�rez, G. Maggio, D. Maino, N. Mandolesi, A. Mangilli, P. G. Martin, E. Mart�nez-Gonz�lez, S. Masi, S. Matarrese, P. Mazzotta, P. McGehee, P. R. Meinhold, A. Melchiorri, L. Mendes, A. Mennella, M. Migliaccio, S. Mitra, M.-A. Miville-Desch�nes, D. Molinari, A. Moneti, L. Montier, G. Morgante, D. Mortlock, A. Moss, D. Munshi, J. A. Murphy, P. Naselsky, F. Nati, P. Natoli, C. B. Netterfield, H. U. N�rgaard-Nielsen, F. Noviello, D. Novikov, I. Novikov, N. Oppermann, C. A. Oxborrow, F. Paci, L. Pagano, F. Pajot, D. Paoletti, F. Pasian, G. Patanchon, O. Perdereau, L. Perotto, F. Perrotta, V. Pettorino, F. Piacentini, M. Piat, E. Pierpaoli, D. Pietrobon, S. Plaszczynski, E. Pointecouteau, G. Polenta, L. Popa, G. W. Pratt, G. Pr�zeau, S. Prunet, J.-L. Puget, J. P. Rachen, R. Rebolo, M. Reinecke, M. Remazeilles, C. Renault, A. Renzi, I. Ristorcelli, G. Rocha, C. Rosset, M. Rossetti, G. Roudier, J. A. Rubi�o-Mart�n, B. Ruiz-Granados, B. Rusholme, M. Sandri, D. Santos, M. Savelainen, G. Savini, D. Scott, M. D. Seiffert, E. P. S. Shellard, M. Shiraishi, L. D. Spencer, V. Stolyarov, R. Stompor, R. Sudiwala, R. Sunyaev, D. Sutton, A.-S. Suur-Uski, J.-F. Sygnet, J. A. Tauber, L. Terenzi, L. Toffolatti, M. Tomasi, M. Tristram, M. Tucci, J. Tuovinen, G. Umana, L. Valenziano, J. Valiviita, B. Van Tent, P. Vielva, F. Villa, L. A. Wade, B. D. Wandelt, I. K. Wehus, D. Yvon, A. Zacchei, A. Zonca,
  [Planck Collaboration],
%  ``Planck 2015 results. XIX. Constraints on primordial magnetic fields,''
  arXiv:1502.01594 [astro-ph.CO].
  %%CITATION = ARXIV:1502.01594;%%
  %36 citations counted in INSPIRE as of 11 Nov 2015


%\cite{Adams:1996cq}
\bibitem{Adams:1996cq}
  J.~A.~Adams, U.~H.~Danielsson, D.~Grasso and H.~Rubinstein,
%  ``Distortion of the acoustic peaks in the CMBR due to a primordial magnetic field,''
  Phys.\ Lett.\ B {\bf 388}, 253 (1996). %[astro-ph/9607043].
  %%CITATION = ASTRO-PH/9607043;%%
  %88 citations counted in INSPIRE as of 12 Nov 2015

  %\cite{Subramanian:1998fn}
\bibitem{Subramanian:1998fn}
  K.~Subramanian and J.~D.~Barrow,
%  ``Microwave background signals from tangled magnetic fields,''
  Phys.\ Rev.\ Lett.\  {\bf 81}, 3575 (1998). %[astro-ph/9803261].
  %%CITATION = ASTRO-PH/9803261;%%
  %130 citations counted in INSPIRE as of 12 Nov 2015

\bibitem{Durrer:1999bk}
  R.~Durrer, P.~G.~Ferreira and T.~Kahniashvili,
%  ``Tensor microwave anisotropies from a stochastic magnetic field,''
  Phys.\ Rev.\ D {\bf 61}, 043001 (2000).
  %[astro-ph/9911040].
  %%CITATION = ASTRO-PH/9911040;%%
  %109 citations counted in INSPIRE as of 09 Nov 2014

 \bibitem{Mack:2001gc}
  A.~Mack, T.~Kahniashvili and A.~Kosowsky,
%  ``Microwave background signatures of a primordial stochastic magnetic field,''
  Phys.\ Rev.\ D {\bf 65}, 123004 (2002).
  %[astro-ph/0105504].
  %%CITATION = ASTRO-PH/0105504;%%
  %138 citations counted in INSPIRE as of 09 Nov 2014

\bibitem{Pogosian:2001np}
  L.~Pogosian, T.~Vachaspati and S.~Winitzki,
%  ``Signatures of kinetic and magnetic helicity in the CMBR,''
  Phys.\ Rev.\ D {\bf 65}, 083502 (2002). %[astro-ph/0112536].
  %%CITATION = ASTRO-PH/0112536;%%
  %51 citations counted in INSPIRE as of 12 Nov 2015

%\cite{Subramanian:2003sh}
\bibitem{Subramanian:2003sh}
  K.~Subramanian, T.~R.~Seshadri and J.~D.~Barrow,
%  ``Small - scale CMB polarization anisotropies due to tangled primordial magnetic fields,''
  Mon.\ Not.\ Roy.\ Astron.\ Soc.\  {\bf 344}, L31 (2003). %[astro-ph/0303014].
  %%CITATION = ASTRO-PH/0303014;%%
  %79 citations counted in INSPIRE as of 12 Nov 2015

%\cite{Subramanian:2002nh}
\bibitem{Subramanian:2002nh}
  K.~Subramanian and J.~D.~Barrow,
%  ``Small-scale microwave background anisotropies due to tangled primordial magnetic fields,''
  Mon.\ Not.\ Roy.\ Astron.\ Soc.\  {\bf 335}, L57 (2002). %[astro-ph/0205312].
  %%CITATION = ASTRO-PH/0205312;%%
  %64 citations counted in INSPIRE as of 12 Nov 2015

%\cite{Caprini:2003vc}
\bibitem{Caprini:2003vc}
  C.~Caprini, R.~Durrer and T.~Kahniashvili,
%  ``The Cosmic microwave background and helical magnetic fields: The Tensor mode,''
  Phys.\ Rev.\ D {\bf 69}, 063006 (2004).
  %[astro-ph/0304556].
  %%CITATION = ASTRO-PH/0304556;%%
  %82 citations counted in INSPIRE as of 09 Nov 2014

%\cite{Lewis:2004ef}
\bibitem{Lewis:2004ef}
  A.~Lewis,
%  ``CMB anisotropies from primordial inhomogeneous magnetic fields,''
  Phys.\ Rev.\ D {\bf 70}, 043011 (2004). %[astro-ph/0406096].
  %%CITATION = ASTRO-PH/0406096;%%
  %124 citations counted in INSPIRE as of 12 Nov 2015

\bibitem{Kahniashvili:2005vu}
  T.~Kahniashvili,
%  ``Effects of cosmological magnetic helicity on the cmb,''
  Astron.\ Nachr.\  {\bf 327}, 414 (2006).
  %[astro-ph/0510151].
  %%CITATION = ASTRO-PH/0510151;%%
  %6 citations counted in INSPIRE as of 10 Jul 2014

\bibitem{Kahniashvili:2005xe}
  T.~Kahniashvili and B.~Ratra,
%  ``Effects of Cosmological Magnetic Helicity on the Cosmic Microwave Background,''
  Phys.\ Rev.\ D {\bf 71}, 103006 (2005).

\bibitem{Kahniashvili:2006hy}
  T.~Kahniashvili and B.~Ratra,
%  ``CMB anisotropies due to cosmological magnetosonic waves,''
  Phys.\ Rev.\ D {\bf 75}, 023002 (2007).
  %[astro-ph/0611247].
  %%CITATION = ASTRO-PH/0611247;%%
  %42 citations counted in INSPIRE as of 09 Nov 2014

%\cite{Kristiansen:2008tx}
\bibitem{Kristiansen:2008tx}
  J.~R.~Kristiansen and P.~G.~Ferreira,
%  ``Constraining primordial magnetic fields with CMB polarization experiments,''
  Phys.\ Rev.\ D {\bf 77}, 123004 (2008). %[arXiv:0803.3210 [astro-ph]].
  %%CITATION = ARXIV:0803.3210;%%
  %25 citations counted in INSPIRE as of 11 Nov 2015


\bibitem{Paoletti:2008ck}
  D.~Paoletti, F.~Finelli and F.~Paci,
%  ``The full contribution of a stochastic background of magnetic fields to CMB anisotropies,''
  Mon.\ Not.\ Roy.\ Astron.\ Soc.\  {\bf 396}, 523 (2009).
  %[arXiv:0811.0230 [astro-ph]].
  %%CITATION = ARXIV:0811.0230;%%
  %52 citations counted in INSPIRE as of 09 Nov 2014

%\cite{Shaw:2009nf}
\bibitem{Shaw:2009nf}
  J.~R.~Shaw and A.~Lewis,
%  ``Massive Neutrinos and Magnetic Fields in the Early Universe,''
  Phys.\ Rev.\ D {\bf 81}, 043517 (2010). %[arXiv:0911.2714 [astro-ph.CO]].
  %%CITATION = ARXIV:0911.2714;%%
  %42 citations counted in INSPIRE as of 02 Oct 2013


%\cite{Yamazaki:2010jw}
\bibitem{Yamazaki:2010jw}
  D.~G.~Yamazaki, K.~Ichiki, T.~Kajino and G.~.J.~Mathews,
%  ``Constraints on the neutrino mass and the primordial magnetic field from the matter density fluctuation parameter $\sigma_8$,''
  Phys.\ Rev.\ D {\bf 81}, 103519 (2010).
%  [arXiv:1005.1638 [astro-ph.CO]].
  %%CITATION = ARXIV:1005.1638;%%
  %10 citations counted in INSPIRE as of 02 Oct 2013



\bibitem{Ichiki:2011ah}
  K.~Ichiki, K.~Takahashi and N.~Sugiyama,
%``Constraint on the primordial vector mode and its magnetic field generation from seven-year Wilkinson Microwave Anisotropy Probe Observations,''
  Phys.\ Rev.\ D {\bf 85}, 043009 (2012).
  %[arXiv:1112.4705 [astro-ph.CO]].
  %%CITATION = ARXIV:1112.4705;%%
  %10 citations counted in INSPIRE as of 09 Nov 2014

%\cite{Yamazaki:2011eu}
\bibitem{Yamazaki:2011eu}
  D.~G.~Yamazaki, K.~Ichiki, T.~Kajino and G.~J.~Mathew,
%  ``Primordial Magnetic Field Effects on the CMB and Large Scale Structure,''
  Adv.\ Astron.\  {\bf 2010}, 586590 (2010). %[arXiv:1112.4922 [astro-ph.CO]].
  %%CITATION = ARXIV:1112.4922;%%
  %10 citations counted in INSPIRE as of 02 Oct 2013

%\cite{Paoletti:2010rx}
\bibitem{Paoletti:2010rx}
  D.~Paoletti and F.~Finelli,
%  ``CMB Constraints on a Stochastic Background of Primordial Magnetic Fields,''
  Phys.\ Rev.\ D {\bf 83}, 123533 (2011). %[arXiv:1005.0148 [astro-ph.CO]].
  %%CITATION = ARXIV:1005.0148;%%
  %39 citations counted in INSPIRE as of 02 Oct 2013

%\cite{Kunze:2010ys}
\bibitem{Kunze:2010ys}
K.~E.~Kunze,
%``CMB anisotropies in the presence of a stochastic magnetic field,''
Phys.\ Rev.\ D {\bf 83}, 023006 (2011). % [arXiv:1007.3163 [astro-ph.CO]].
%%CITATION = ARXIV:1007.3163;%%


%\cite{Kunze:2011bp}
\bibitem{Kunze:2011bp}
  K.~E.~Kunze,
%  ``Effects of helical magnetic fields on the cosmic microwave background,''
  Phys.\ Rev.\ D {\bf 85}, 083004 (2012). %[arXiv:1112.4797 [astro-ph.CO]].
  %%CITATION = ARXIV:1112.4797;%%
  %15 citations counted in INSPIRE as of 11 Nov 2015

%\cite{Trivedi:2011vt}
\bibitem{Trivedi:2011vt}
  P.~Trivedi, T.~R.~Seshadri and K.~Subramanian,
%  ``Cosmic Microwave Background Trispectrum and Primordial Magnetic Field Limits,''
  Phys.\ Rev.\ Lett.\  {\bf 108}, 231301 (2012).%[arXiv:1111.0744 [astro-ph.CO]].
  %%CITATION = ARXIV:1111.0744;%%
  %21 citations counted in INSPIRE as of 11 Nov 2015

\bibitem{Paoletti:2012bb}
  D.~Paoletti and F.~Finelli,
%  ``Constraints on a Stochastic Background of Primordial Magnetic Fields with WMAP and South Pole Telescope data,''
  Phys.\ Lett.\ B {\bf 726}, 45 (2013).
  %[arXiv:1208.2625 [astro-ph.CO]].
  %%CITATION = ARXIV:1208.2625;%%
  %26 citations counted in INSPIRE as of 08 Nov 2014

\bibitem{Chen:2013gva}
  P.~Chen and T.~Suyama,
%  ``Constraining Primordial Magnetic Fields by CMB Photon-Graviton Conversion,''
  Phys.\ Rev.\ D {\bf 88}, 123521 (2013).
  %[arXiv:1309.0537 [astro-ph.CO]].
  %%CITATION = ARXIV:1309.0537;%%
  %2 citations counted in INSPIRE as of 08 Nov 2014


\bibitem{Yamazaki:2013hda}
  D.~G.~Yamazaki, K.~Ichiki and K.~Takahashi,
%  ``Constraints on the multi-lognormal magnetic fields from the observations of the cosmic microwave background and the matter power spectrum,''
  Phys.\ Rev.\ D {\bf 88}, no. 10, 103011 (2013).
  %[arXiv:1311.2584 [astro-ph.CO]].
  %%CITATION = ARXIV:1311.2584;%%
  %2 citations counted in INSPIRE as of 08 Nov 2014


\bibitem{Trivedi:2013wqa}
  P.~Trivedi, K.~Subramanian and T.~R.~Seshadri,
%  ``Primordial Magnetic Field Limits from CMB Trispectrum---Scalar Modes and Planck Constraints,''
  Phys.\ Rev.\ D {\bf 89}, 043523 (2014).
  %[arXiv:1312.5308 [astro-ph.CO]].
  %%CITATION = ARXIV:1312.5308;%%
  %6 citations counted in INSPIRE as of 08 Nov 2014


%\cite{Kahniashvili:2014dfa}
\bibitem{Kahniashvili:2014dfa}
  T.~Kahniashvili, Y.~Maravin, G.~Lavrelashvili and A.~Kosowsky,
%  ``Primordial Magnetic Helicity Constraints from WMAP Nine-Year Data,''
  Phys.\ Rev.\ D {\bf 90}, 083004 (2014). %[arXiv:1408.0351 [astro-ph.CO]].
  %%CITATION = ARXIV:1408.0351;%%

%\cite{Ballardini:2014jta}
\bibitem{Ballardini:2014jta}
  M.~Ballardini, F.~Finelli and D.~Paoletti,
%  ``CMB anisotropies generated by a stochastic background of primordial magnetic fields with non-zero helicity,''
  JCAP {\bf 1510}, no. 10, 031 (2015). %[arXiv:1412.1836 [astro-ph.CO]].
  %%CITATION = ARXIV:1412.1836;%%
  %2 citations counted in INSPIRE as of 11 Nov 2015
%%%%%%%%%%%%%%%%%%%%%%%%%%%%%%%%%%%%%%%%%%%
%BROKEN ISOTROPY




%%%%%%%%%%%%%%%%%%%%%%%%%%%%%%%%%%%%%%%%%%%%%%%%%%%%%%%%%%%%


%%%%%%%%%%%%%%%%%%%%%%%%%%%%%%%%%%%%%%%%%%%
%DISTORTIONS
\bibitem{Jedamzik:1999bm}
  K.~Jedamzik, V.~Katalinic and A.~V.~Olinto,
%  ``A Limit on primordial small scale magnetic fields from CMB distortions,''
  Phys.\ Rev.\ Lett.\  {\bf 85}, 700 (2000). %[astro-ph/9911100].
  %%CITATION = ASTRO-PH/9911100;%%
  %106 citations counted in INSPIRE as of 11 Nov 2015

%\cite{Dent:2012ne}
\bibitem{Dent:2012ne}
  J.~B.~Dent, D.~A.~Easson and H.~Tashiro,
%  ``Cosmological constraints from CMB distortion,''
  Phys.\ Rev.\ D {\bf 86}, 023514 (2012).%[arXiv:1202.6066 [astro-ph.CO]].
  %%CITATION = ARXIV:1202.6066;%%
  %32 citations counted in INSPIRE as of 11 Nov 2015

\bibitem{Tashiro:2013yea}
  H.~Tashiro, J.~Silk and D.~J.~E.~Marsh,
%  ``Constraints on primordial magnetic fields from CMB distortions in the axiverse,''
  Phys.\ Rev.\ D {\bf 88}, 125024 (2013).
  [arXiv:1308.0314 [astro-ph.CO]].
  %%CITATION = ARXIV:1308.0314;%%
  %10 citations counted in INSPIRE as of 08 Nov 2014


%\cite{Kunze:2013uja}
\bibitem{Kunze:2013uja}
  K.~E.~Kunze and E.~Komatsu,
%  ``Constraining primordial magnetic fields with distortions of the black-body spectrum of the cosmic microwave background: pre- and post-decoupling contributions,''
  JCAP {\bf 1401}, 009 (2014). %[arXiv:1309.7994 [astro-ph.CO]].
  %%CITATION = ARXIV:1309.7994;%%
  %19 citations counted in INSPIRE as of 11 Nov 2015

%\cite{Miyamoto:2013oua}
\bibitem{Miyamoto:2013oua}
  K.~Miyamoto, T.~Sekiguchi, H.~Tashiro and S.~Yokoyama,
%  ``CMB distortion anisotropies due to the decay of primordial magnetic fields,''
  Phys.\ Rev.\ D {\bf 89}, no. 6, 063508 (2014). %[arXiv:1310.3886 [astro-ph.CO]].
  %%CITATION = ARXIV:1310.3886;%%
  %9 citations counted in INSPIRE as of 11 Nov 2015

%\cite{Jedamzik:2013gua}
\bibitem{Jedamzik:2013gua}
  K.~Jedamzik and T.~Abel,
%  ``Small-scale primordial magnetic fields and anisotropies in the cosmic microwave background radiation,''
  JCAP {\bf 1310}, 050 (2013).
  %%CITATION = JCAPA,1310,050;%%
  %2 citations counted in INSPIRE as of 11 Nov 2015

%\cite{Amin:2014ada}
\bibitem{Amin:2014ada}
  M.~A.~Amin and D.~Grin,
%  ``Probing early-universe phase transitions with CMB spectral distortions,''
  Phys.\ Rev.\ D {\bf 90}, no. 8, 083529 (2014). %[arXiv:1405.1039 [astro-ph.CO]].
  %%CITATION = ARXIV:1405.1039;%%
  %5 citations counted in INSPIRE as of 11 Nov 2015

\bibitem{Kunze:2014eka}
  K.~E.~Kunze and E.~Komatsu,
%  ``Constraints on primordial magnetic fields from the optical depth of the cosmic microwave background,''
  JCAP {\bf 1506},  027 (2015). %[arXiv:1501.00142 [astro-ph.CO]].
  %%CITATION = ARXIV:1501.00142;%%
  %2 citations counted in INSPIRE as of 11 Nov 2015

%\cite{Chluba:2015lpa}
\bibitem{Chluba:2015lpa}
  J.~Chluba, D.~Paoletti, F.~Finelli and J.~A.~Rubi�o-Mart�n,
%  ``Effect of primordial magnetic fields on the ionization history,''
  Mon.\ Not.\ Roy.\ Astron.\ Soc.\  {\bf 451}, no. 2, 2244 (2015). %[arXiv:1503.04827 [astro-ph.CO]].
  %%CITATION = ARXIV:1503.04827;%%
  %4 citations counted in INSPIRE as of 11 Nov 2015

%\cite{Wagstaff:2015jaa}
\bibitem{Wagstaff:2015jaa}
  J.~M.~Wagstaff and R.~Banerjee,
%  ``CMB spectral distortions from the decay of causally generated magnetic fields,''
  arXiv:1508.01683 [astro-ph.CO].
  %%CITATION = ARXIV:1508.01683;%%
%%%%%%%%%%%%%%%%%%%%%%%%%%%%%%%%%%%%%%%%%%%%
%%%%%%%%%%%%%%%%%%%%%%%%%%%%%%%%%%%%%%%%

%\cite{Durrer:1998ya}
\bibitem{Durrer:1998ya}
  R.~Durrer, T.~Kahniashvili and A.~Yates,
%  ``Microwave background anisotropies from Alfven waves,''
  Phys.\ Rev.\ D {\bf 58}, 123004 (1998).
  %[astro-ph/9807089].
  %%CITATION = ASTRO-PH/9807089;%%


\bibitem{Chen:2004nf}
  G.\ Chen,  P. Mukherjee, T. Kahniashvili, B. Ratra, Y. Wang,
%  "Looking for Cosmological Alfv\'{e}n Waves in Wilkinson Microwave Anisotropy Probe Data''
  Astrophys.\ J.\ {\bf 611}, 655 (2004).

  \bibitem{Brown:2005}
  I.\ Brown and R.\ Crittenden,
%  ``Non-Gaussianity from cosmic magnetic fields,'',
  Phys.\ Rev.\  D {\bf 72}, 063002 (2005).

  \bibitem{Demianski:2007}
  M.\ Demianski and A.\ G.\ Doroshkevich,
%  ``Extension of the standard cosmological model: Anisotropy, rotation, and magnetic field,'',
 Phys.\ Rev.\  D
      {\bf 75}, 123517 (2007).

  \bibitem{Bernui:2008}
  A.\ Bernui and W.\ S.\ Hipolito-Ricaldi,
%  ``Can a primordial magnetic field originate large-scale anomalies in WMAP data?,'',
  Mon.\ Not.\ R.\
      Astron.\ Soc.\ {\bf 389}, 1453 (2008).

 \bibitem{Samal:2009}
  P.\ K.\ Samal, R.\ Saha, P.\ Jain and J.\ P.\ Ralston,
%  ``Testing Isotropy of Cosmic Microwave Background Radiation,'',
      Mon.\ Not.\ R.\ Astron.\ Soc.\ {\bf 396}, 511 (2009).

\bibitem{Kahniashvili:2008sh}
  T.~Kahniashvili, G.~Lavrelashvili and B.~Ratra,
%  ``CMB Temperature Anisotropy from Broken Spatial Isotropy due to an Homogeneous Cosmological Magnetic Field,''
  Phys.\ Rev.\ D {\bf 78}, 063012 (2008).
  %[arXiv:0807.4239 [astro-ph]].
  %%CITATION = ARXIV:0807.4239;%%
  %61 citations counted in INSPIRE as of 09 Nov 2014


%\cite{Naselsky:2008ei}
\bibitem{Naselsky:2008ei}
  P.~Naselsky and J.~Kim,
%  ``Fast simulation of the whole-sky CMB map in the presence of primordial magnetic field,''
    arXiv:0804.3467 [astro-ph].
  %%CITATION = ARXIV:0804.3467;%%
  %5 citations counted in INSPIRE as of 11 Nov 201

%\cite{Kim:2009gi}
\bibitem{Kim:2009gi}
  J.~Kim and P.~Naselsky,
%  ``Alfv\'en turbulence in the WMAP 5 year data and a forecast for the PLANCK,''
  JCAP {\bf 0907}, 041 (2009). %[arXiv:0903.1930 [astro-ph.CO]].
  %%CITATION = ARXIV:0903.1930;%%
  %25 citations counted in INSPIRE as of 11 Nov 2015

%\cite{Ade:2013nlj}
\bibitem{Ade:2013nlj}
P.~A.~R.~Ade,
%N. Aghanim, C. Armitage-Caplan, M. Arnaud, M. Ashdown, F. Atrio-Barandela, J. Aumont, C. Baccigalupi86, A. J. Banday, R. B. Barreiro, J. G. Bartlett1, N. Bartolo, E. Battaner, R. Battye, K. Benabed, A. Beno, A. Benoit-Levy, J.-P. Bernard, M. Bersanelli, P. Bielewicz, J. Bobin, J. J. Bock, A. Bonaldi, L. Bonavera, J. R. Bond, J. Borrill13, F. R. Bouchet, M. Bridges, M. Bucher1, C. Burigana, R. C. Butler, J.-F. Cardoso74;1;61, A. Catalano, A. Challinor, A. Chamballu, R.-R. Chary, L.-Y Chiang, H. C. Chiang, P. R. Christensen, S. Church, D. L. Clements, S. Colombi, L. P. L. Colombo, F. Couchot, A. Coulais, B. P. Crill, M. Cruz, A. Curto, F. Cuttaia, L. Danese, R. D. Davies, R. J. Davis, P. de Bernardis, A. de Rosa, G. de Zotti, J. Delabrouille, J.-M. Delouis, F.-X. Desert, J. M. Diego, H. Dole, S. Donzelli, O. Dor�, M. Douspis, A. Ducout, X. Dupac, G. Efstathiou, F. Elsner, T. A. En�lin, H. K. Eriksen, Y. Fantaye, J. Fergusson, F. Finelli, O. Forni, M. Frailis, E. Franceschi, M. Frommert, S. Galeotta, K. Ganga, M. Giard, G. Giardino, Y. Giraud-H�raud, J. Gonz�lez-Nuevo, K. M. G�rski, S. Gratton, A. Gregorio, A. Gruppuso, F. K. Hansen, M. Hansen, D. Hanson, D. L. Harrison, G. Helou, S. Henrot-Versill�, C. Hern�ndez-Monteagudo, D. Herranz, S. R. Hildebrandt, E. Hivon, M. Hobson, W. A. Holmes, A. Hornstrup, W. Hovest, K. M. Huffenberger, A. H. Jaffe, T. R. Jaffe, W. C. Jones, M. Juvela, E. Keih�nen, R. Keskitalo, J. Kim, T. S. Kisner, J. Knoche, L. Knox, M. Kunz, H. Kurki-Suonio, G. Lagache, A. L�hteenm�ki, J.-M. Lamarre, A. Lasenby, R. J. Laureijs, C. R. Lawrence, J. P. Leahy, R. Leonardi, C. Leroy, J. Lesgourgues, M. Liguori, P. B. Lilje, M. Linden-V�rnle, M. L�pez-Caniego, P. M. Lubin, J. F. Mac�as-P�rez, B. Maffei, D. Maino, N. Mandolesi, A. Mangilli, D. Marinucci, M. Maris, D. J. Marshall, P. G. Martin, E. Mart�nez-Gonz�lez, S. Masi, M. Massardi, S. Matarrese, F. Matthai, P. Mazzotta, J. D. McEwen, P. R. Meinhold, A. Melchiorri, L. Mendes, A. Mennella, M. Migliaccio, K. Mikkelsen, S. Mitra, M.-A. Miville-Desch�nes, D. Molinari, A. Moneti, L. Montier, G. Morgante, D. Mortlock, A. Moss, D. Munshi, J. A. Murphy, P. Naselsky, F. Nati, P. Natoli, C. B. Netterfield, H. U. N�rgaard-Nielsen, F. Noviello, D. Novikov, I. Novikov, S. Osborne, C. A. Oxborrow, F. Paci, L. Pagano, F. Pajot, D. Paoletti, F. Pasian, G. Patanchon, H. V. Peiris, O. Perdereau, L. Perotto, F. Perrotta, F. Piacentini, M. Piat, E. Pierpaoli, D. Pietrobon, S. Plaszczynski, D. Pogosyan, E. Pointecouteau, G. Polenta, N. Ponthieu, L. Popa, T. Poutanen, G. W. Pratt, G. Pr�zeau, S. Prunet, J.-L. Puget, J. P. Rachen, B. Racine, C. R�th, R. Rebolo, M. Reinecke, M. Remazeilles, C. Renault, A. Renzi, S. Ricciardi, T. Riller, I. Ristorcelli, G. Rocha, C. Rosset, A. Rotti, G. Roudier, J. A. Rubi�o-Mart�n, B. Ruiz-Granados, B. Rusholme, M. Sandri, D. Santos, G. Savini, D. Scott, M. D. Seiffert, E. P. S. Shellard, T. Souradeep, L. D. Spencer, J.-L. Starck, V. Stolyarov, R. Stompor, R. Sudiwala, F. Sureau, P. Sutter, D. Sutton, A.-S. Suur-Uski, J.-F. Sygnet, J. A. Tauber, D. Tavagnacco, L. Terenzi, L. Toffolatti, M. Tomasi, M. Tristram, M. Tucci, J. Tuovinen, M. T�rler, L. Valenziano, J. Valiviita, B. Van Tent, J. Varis, P. Vielva, F. Villa, N. Vittorio, L. A. Wade, B. D. Wandelt, I. K. Wehus, M. White, A. Wilkinson, D. Yvon, A. Zacchei, A. Zonca,
{\it et al.}
[Planck Collaboration],
%``Planck 2013 results. XXIII. Isotropy and Statistics of the CMB,''
Astron. Astrophys.  {\bf 571}, A23, (2014).
%arXiv:1303.5083 [astro-ph.CO].
%%CITATION = ARXIV:1303.5083;%%



%%%%%%%%%%%%%%%%%%%%%%%%%%%%%%%%%%%%%%
%BBN
\bibitem{Kawasaki:2012va}
  M.~Kawasaki and M.~Kusakabe,
%  ``Updated constraint on a primordial magnetic field during big bang nucleosynthesis and a formulation of field effects,''
  Phys.\ Rev.\ D {\bf 86}, 063003 (2012).
  %[arXiv:1204.6164 [astro-ph.CO]].
  %%CITATION = ARXIV:1204.6164;%%
  %14 citations counted in INSPIRE as of 08 Nov 2014

\bibitem{Yamazaki:2014fja}
  D.~G.~Yamazaki, M.~Kusakabe, T.~Kajino, G.~J.~Mathews and M.~K.~Cheoun,
%  ``Cosmological solutions to the Lithium problem: Big-bang nucleosynthesis with photon cooling, $X$-particle decay and a primordial magnetic field,''
  Phys.\ Rev.\ D {\bf 90}, 023001 (2014).
  %[arXiv:1407.0021 [astro-ph.CO]].
  %%CITATION = ARXIV:1407.0021;%%

\bibitem{Yamazaki:2012jd}
  D.~G.~Yamazaki and M.~Kusakabe,
%  ``Effects of power law primordial magnetic field on big bang nucleosynthesis,''
  Phys.\ Rev.\ D {\bf 86}, 123006 (2012).
  %[arXiv:1212.2968 [astro-ph.CO]].
  %%CITATION = ARXIV:1212.2968;%%
  %4 citations counted in INSPIRE as of 08 Nov 2014

\bibitem{Widrow:2011hs}
  L.~M.~Widrow, D.~Ryu, D.~R.~G.~Schleicher, K.~Subramanian, C.~G.~Tsagas and R.~A.~Treumann,
%  ``The First Magnetic Fields,''
  Space Sci.\ Rev.\  {\bf 166}, 37 (2012).
  %[arXiv:1109.4052 [astro-ph.CO]].
  %%CITATION = ARXIV:1109.4052;%%
  %45 citations counted in INSPIRE as of 08 Nov 2014

\bibitem{Shaw:2010ea}
J.~R.~Shaw and A.~Lewis,
%  ``Constraining Primordial Magnetism,''
  Phys.\ Rev.\ D {\bf 86}, 043510 (2012).
  %arXiv:1006.4242 [astro-ph.CO],

%%%%%%%%%%%%%%%%%%%%%%%%%%%%%%%%%%%%%%%%%%%%%%%%%%%%%%%%%%%%%%%%%%%%%%%%%%%%%%%%%%%%%
\bibitem{magfields2}
T. G. Cowling,
%"The dissipation of magnetic energy in an ionized gas",
Mon. Not. Roy. Astron. Soc., 116, 114 (1956).

\bibitem{magfields3}
L. Mestel and L. Spitzer,
%"Star formation in magnetic dust clouds",
Mon. Not. Roy. Astron. Soc., 116, 503 (1956).

\bibitem{Jedamzik:1996wp}
  K.~Jedamzik, V.~Katalinic and A.~V.~Olinto,
%  ``Damping of cosmic magnetic fields,''
  Phys.\ Rev.\ D {\bf 57}, 3264 (1998). %[astro-ph/9606080].
  %%CITATION = ASTRO-PH/9606080;%%
  %185 citations counted in INSPIRE as of 12 Nov 2015

\bibitem{Sethi:2004pe}
  S.~K.~Sethi and K.~Subramanian,
%  ``Primordial magnetic fields in the post-recombination era and early reionization,''
  Mon.\ Not.\ Roy.\ Astron.\ Soc.\  {\bf 356}, 778 (2005).
  %[astro-ph/0405413].
  %%CITATION = ASTRO-PH/0405413;%%
  %49 citations counted in INSPIRE as of 08 Nov 2014

%\cite{Tashiro:2005ua}
\bibitem{Tashiro:2005ua}
  H.~Tashiro and N.~Sugiyama,
%  ``The early reionization with the primordial magnetic fields,''
  Mon.\ Not.\ Roy.\ Astron.\ Soc.\  {\bf 368}, 965 (2006). %[astro-ph/0512626].
  %%CITATION = ASTRO-PH/0512626;%%
  %35 citations counted in INSPIRE as of 12 Nov 2015

%\cite{Tashiro:2005hc}
\bibitem{Tashiro:2005hc}
  H.~Tashiro, N.~Sugiyama and R.~Banerjee,
%  ``Nonlinear evolution of cosmic magnetic fields and cosmic microwave background anisotropies,''
  Phys.\ Rev.\ D {\bf 73}, 023002 (2006). %[astro-ph/0509220].
  %%CITATION = ASTRO-PH/0509220;%%
  %25 citations counted in INSPIRE as of 12 Nov 2015

\bibitem{Tashiro:2006uv}
  H.~Tashiro and N.~Sugiyama,
%  ``Probing Primordial Magnetic Fields with the 21cm Fluctuations,''
  Mon.\ Not.\ Roy.\ Astron.\ Soc.\  {\bf 372}, 1060 (2006).
  %[astro-ph/0607169].
  %%CITATION = ASTRO-PH/0607169;%%
  %18 citations counted in INSPIRE as of 08 Nov 2014


%\cite{Schleicher:2008aa}
\bibitem{Schleicher:2008aa}
  D.~R.~G.~Schleicher, R.~Banerjee and R.~S.~Klesser,
%  ``Reionization - A probe for the stellar population and the physics of the early universe,''
  Phys.\ Rev.\ D {\bf 78}, 083005 (2008). %[arXiv:0807.3802 [astro-ph]].
  %%CITATION = ARXIV:0807.3802;%%
  %41 citations counted in INSPIRE as of 11 Nov 2015



\bibitem{Schleicher:2008hc}
  D.~R.~G.~Schleicher, R.~Banerjee and R.~S.~Klessen,
%  ``Influence of primordial magnetic fields on 21 cm emission,''
  Astrophys.\ J.\  {\bf 692}, 236 (2009).
  %[arXiv:0808.1461 [astro-ph]].
  %%CITATION = ARXIV:0808.1461;%%
  %16 citations counted in INSPIRE as of 08 Nov 2014


%\cite{Sethi:2009dd}
\bibitem{Sethi:2009dd}
  S.~K.~Sethi and K.~Subramanian,
%  ``Primordial magnetic fields and the HI signal from the epoch of reionization,''
  JCAP {\bf 0911}, 021 (2009). %[arXiv:0911.0244 [astro-ph.CO]].
  %%CITATION = ARXIV:0911.0244;%%
  %17 citations counted in INSPIRE as of 12 Nov 2015

\bibitem{Tashiro:2009hx}
  H.~Tashiro and N.~Sugiyama,
%  ``S-Z power spectrum produced by primordial magnetic fields,''
  Mon. Not. Roy. Astron. Soc. {\bf 411 }, 1284 (2011).
  %arXiv:0908.0113 [astro-ph.CO].
  %%CITATION = ARXIV:0908.0113;%%
  %5 citations counted in INSPIRE as of 08 Nov 2014


%\cite{Sethi:2008eq}
\bibitem{Sethi:2008eq}
  S.~K.~Sethi, B.~B.~Nath and K.~Subramanian,
%  ``Primordial magnetic fields and formation of molecular hydrogen,''
  Mon.\ Not.\ Roy.\ Astron.\ Soc.\  {\bf 387}, 1589 (2008). %[arXiv:0804.3473 [astro-ph]].
  %%CITATION = ARXIV:0804.3473;%%
  %24 citations counted in INSPIRE as of 12 Nov 2015



%\cite{Yamazaki:2010nf}
\bibitem{Yamazaki:2010nf}
  D.~G.~Yamazaki, K.~Ichiki, T.~Kajino and G.~J.~Mathews,
%  ``New Constraints on the Primordial Magnetic Field,''
  Phys.\ Rev.\ D {\bf 81}, 023008 (2010).% [arXiv:1001.2012 [astro-ph.CO]].
  %%CITATION = ARXIV:1001.2012;%%
  %36 citations counted in INSPIRE as of 02 Oct 2013

%\cite{Schleicher:2011jj}
\bibitem{Schleicher:2011jj}
  D.~R.~G.~Schleicher and F.~Miniati,
%  ``Primordial magnetic field constraints from the end of reionization,''
  Mon. Not. Roy. Astron. Soc. Lett., {\bf 418}, L143 (2011).
  %arXiv:1108.1874 [astro-ph.CO].  %%CITATION = ARXIV:1108.1874;%%


\bibitem{Kahniashvili:2012dy}
  T.~Kahniashvili, Y.~Maravin, A.~Natarajan, N.~Battaglia and A.~G.~Tevzadze,
%  ``Constraining primordial magnetic fields through large scale structure,''
  Astrophys.\ J.\  {\bf 770}, 47 (2013).




\bibitem{Fedeli:2012rr}
  C.~Fedeli and L.~Moscardini,
%  ``Constraining Primordial Magnetic Fields with Future Cosmic Shear Surveys,''
  JCAP {\bf 1211}, 055 (2012).
  %[arXiv:1209.6332 [astro-ph.CO]].
  %%CITATION = ARXIV:1209.6332;%%
  %3 citations counted in INSPIRE as of 08 Nov 2014

\bibitem{Pandey:2012ss}
  K.~L.~Pandey and S.~K.~Sethi,
%  ``Probing Primordial Magnetic Fields Using Ly-alpha Clouds,''
  Astrophys.\ J.\  {\bf 762}, 15 (2013).
  %[arXiv:1210.3298 [astro-ph.CO]].
  %%CITATION = ARXIV:1210.3298;%%
  %14 citations counted in INSPIRE as of 08 Nov 2014

\bibitem{Pandey:2014vga}
  K.~L.~Pandey, T.~R.~Choudhury, S.~K.~Sethi and A.~Ferrara,
%  ``Reionization constraints on primordial magnetic fields,''
  Mon.\ Not.\ Roy.\ Astron.\ Soc.\  {\bf 451}, no. 2, 1692 (2015). %[arXiv:1410.0368 [astro-ph.CO]].
  %%CITATION = ARXIV:1410.0368;%%
  %1 citations counted in INSPIRE as of 12 Nov 2015


%\cite{Venumadhav:2014tqa}
\bibitem{Venumadhav:2014tqa}
  T.~Venumadhav, A.~Oklopcic, V.~Gluscevic, A.~Mishra and C.~M.~Hirata,
%  ``A new probe of magnetic fields in the pre-reionization epoch: I. Formalism,''
  arXiv:1410.2250 [astro-ph.CO].
  %%CITATION = ARXIV:1410.2250;%%

\bibitem{Vasiliev:2014vpa}
  E.~O.~Vasiliev and S.~K.~Sethi,
%  ``H I Absorption from the Epoch of Reionization and Primordial Magnetic Fields,''
  Astrophys.\ J.\  {\bf 786}, 142 (2014).
  %doi:10.1088/0004-637X/786/2/142
  %[arXiv:1403.5650 [astro-ph.CO]].
  %%CITATION = doi:10.1088/0004-637X/786/2/142;%%
  %1 citations counted in INSPIRE as of 16 Nov 2015

%\cite{Ade:2015kva}
\bibitem{Ade:2015kva}
  P. A. R. Ade,
%N. Aghanim, M. I. R. Alves, M. Arnaud, D. Arzoumanian, M. Ashdown, J. Aumont, C. Baccigalupi, A. J. Banday, R. B. Barreiro, N. Bartolo, E. Battaner, K. Benabed, A. Beno�t, A. Benoit-L�vy, J.-P. Bernard, M. Bersanelli, P. Bielewicz, J. J. Bock, L. Bonavera, J. R. Bond, J. Borrill, F. R. Bouchet, F. Boulanger, A. Bracco, C. Burigana, E. Calabrese, J.-F. Cardoso, A. Catalano, H. C. Chiang, P. R. Christensen, L. P. L. Colombo, C. Combet, F. Couchot, B. P. Crill, A. Curto, F. Cuttaia, L. Danese, R. D. Davies, R. J. Davis, P. de Bernardis, A. de Rosa, G. de Zotti, J. Delabrouille, C. Dickinson, J. M. Diego, H. Dole, S. Donzelli, O. Dor�, M. Douspis, A. Ducout, X. Dupac, G. Efstathiou, F. Elsner, T. A. En�lin, H. K. Eriksen, D. Falceta-Gon�alves, E. Falgarone, K. Ferri�re, F. Finelli, O. Forni, M. Frailis, A. A. Fraisse, E. Franceschi, A. Frejsel, S. Galeotta, S. Galli, K. Ganga, T. Ghosh, M. Giard, E. Gjerl�w, J. Gonz�lez-Nuevo, K. M. G�rski, A. Gregorio, A. Gruppuso, J. E. Gudmundsson, V. Guillet, D. L. Harrison, G. Helou, P. Hennebelle, S. Henrot-Versill�, C. Hern�ndez-Monteagudo, D. Herranz, S. R. Hildebrandt, E. Hivon, W. A. Holmes, A. Hornstrup, K. M. Huffenberger, G. Hurier, A. H. Jaffe, T. R. Jaffe, W. C. Jones, M. Juvela, E. Keih�nen, R. Keskitalo, T. S. Kisner, J. Knoche, M. Kunz, H. Kurki-Suonio, G. Lagache, J.-M. Lamarre, A. Lasenby, M. Lattanzi, C. R. Lawrence, R. Leonardi, F. Levrier, M. Liguori, P. B. Lilje, M. Linden-V�rnle, M. L�pez-Caniego, P. M. Lubin, J. F. Mac�as-P�rez, D. Maino, N. Mandolesi, A. Mangilli, M. Maris, P. G. Martin, E. Mart�nez-Gonz�lez, S. Masi, S. Matarrese, A. Melchiorri, L. Mendes, A. Mennella, M. Migliaccio, M.-A. Miville-Desch�nes, A. Moneti, L. Montier, G. Morgante, D. Mortlock, D. Munshi, J. A. Murphy, P. Naselsky, F. Nati, C. B. Netterfield, F. Noviello, D. Novikov, I. Novikov, N. Oppermann, C. A. Oxborrow, L. Pagano, F. Pajot, R. Paladini, D. Paoletti, F. Pasian, L. Perotto, V. Pettorino, F. Piacentini, M. Piat, E. Pierpaoli, D. Pietrobon, S. Plaszczynski, E. Pointecouteau, G. Polenta, N. Ponthieu, G. W. Pratt, S. Prunet, J.-L. Puget, J. P. Rachen, M. Reinecke, M. Remazeilles, C. Renault, A. Renzi, I. Ristorcelli, G. Rocha, M. Rossetti, G. Roudier, J. A. Rubi�o-Mart�n, B. Rusholme, M. Sandri, D. Santos, M. Savelainen, G. Savini, D. Scott, J. D. Soler, V. Stolyarov, R. Sudiwala, D. Sutton, A.-S. Suur-Uski, J.-F. Sygnet, J. A. Tauber, L. Terenzi, L. Toffolatti, M. Tomasi, M. Tristram, M. Tucci, G. Umana, L. Valenziano, J. Valiviita, B. Van Tent, P. Vielva, F. Villa, L. A. Wade, B. D. Wandelt, I. K. Wehus, N. Ysard, D. Yvon, A. Zonca,
  {\it et al.}
  [Planck Collaboration],
 % ``Planck intermediate results. XXXV. Probing the role of the magnetic field in the formation of structure in molecular clouds,''
  arXiv:1502.04123 [astro-ph.GA].
  %%CITATION = ARXIV:1502.04123;%%
  %2 citations counted in INSPIRE as of 12 Nov 2015

\bibitem{Miniati:2010ne}
F. Miniati and  A. R.  Bell,
%``Resistive Magnetic Field Generation at Cosmic Dawn,''
Astrophys.\ J.\  {\bf 729}, 73 (2011).
% [arXiv:1001.2011 [astro-ph.CO]].  %%CITATION = ARXIV:1001.2011;%%


%%%%%%%%%%%%%%%%%%%%%%%%%%%%%%%%%%%%%%%%%%%%%%%%%%%%%%%%%%%%%%%%%%%%%%%%%%%%%
\bibitem{Kahniashvili:2009qi}
  T.~Kahniashvili, A.~G.~Tevzadze and B.~Ratra,
%  ``Phase Transition Generated Cosmological Magnetic Field at Large Scales,''
  Astrophys.\ J.\  {\bf 726}, 78 (2011).
  %[arXiv:0907.0197 [astro-ph.CO]].
  %%CITATION = ARXIV:0907.0197;%%
  %24 citations counted in INSPIRE as of 08 Nov 2014

%\cite{Kahniashvili:2010gp}
\bibitem{Kahniashvili:2010gp}
  T.~Kahniashvili, A.~Brandenburg, A.~G.~Tevzadze and B.~Ratra,
%  ``Numerical simulations of the decay of primordial magnetic turbulence,''
  Phys.\ Rev.\ D {\bf 81}, 123002 (2010). %[arXiv:1004.3084 [astro-ph.CO]].
  %%CITATION = ARXIV:1004.3084;%%
  %21 citations counted in INSPIRE as of 12 Nov 2015

%\cite{Kahniashvili:2012vt}
\bibitem{Kahniashvili:2012vt}
  T.~Kahniashvili, A.~Brandenburg, L.~Campanelli, B.~Ratra and A.~G.~Tevzadze,
%  ``Evolution of inflation-generated magnetic field through phase transitions,''
  Phys.\ Rev.\ D {\bf 86}, 103005 (2012).%[arXiv:1206.2428 [astro-ph.CO]].
  %%CITATION = ARXIV:1206.2428;%%
  %20 citations counted in INSPIRE as of 12 Nov 2015

%\cite{Tevzadze:2012kk}
\bibitem{Tevzadze:2012kk}
  A.~G.~Tevzadze, L.~Kisslinger, A.~Brandenburg and T.~Kahniashvili,
%  ``Magnetic Fields from QCD Phase Transitions,''
  Astrophys.\ J.\  {\bf 759}, 54 (2012).%[arXiv:1207.0751 [astro-ph.CO]].
  %%CITATION = ARXIV:1207.0751;%%
  %18 citations counted in INSPIRE as of 12 Nov 2015

%\cite{Kahniashvili:2012uj}
\bibitem{Kahniashvili:2012uj}
  T.~Kahniashvili, A.~G.~Tevzadze, A.~Brandenburg and A.~Neronov,
%  ``Evolution of Primordial Magnetic Fields from Phase Transitions,''
  Phys.\ Rev.\ D {\bf 87}, no. 8, 083007 (2013). %[arXiv:1212.0596 [astro-ph.CO]].
  %%CITATION = ARXIV:1212.0596;%%
  %28 citations counted in INSPIRE as of 12 Nov 2015

%\cite{Brandenburg:2014mwa}
\bibitem{Brandenburg:2014mwa}
  A.~Brandenburg, T.~Kahniashvili and A.~G.~Tevzadze,
%  ``Nonhelical inverse transfer of a decaying turbulent magnetic field,''
  Phys.\ Rev.\ Lett.\  {\bf 114}, no. 7, 075001 (2015). %[arXiv:1404.2238 [astro-ph.CO]].
  %%CITATION = ARXIV:1404.2238;%%
  %17 citations counted in INSPIRE as of 12 Nov 2015

%\cite{Kahniashvili:2015msa}
\bibitem{Kahniashvili:2015msa}
  T.~Kahniashvili, A.~Brandenburg and A.~G.~Tevzadze,
%  ``Evolution of Primordial Magnetic Fields: From Generation Till Today,''
  arXiv:1507.00510 [astro-ph.CO].
  %%CITATION = ARXIV:1507.00510;%%
  %1 citations counted in INSPIRE as of 12 Nov 2015

\bibitem{kbtv15} A. Kahniashvili, A. Brandenburg, A. Tevzadze, T. Vachaspati,
``Helical Magnetic Fields from Phase Transitions'',
2016, in preparation.

\bibitem{kbdty15}
A. Kahniashvili, R. Durrer, A. Brandenburg,  A. Tevzadze, W. Yin,
``Inflationary Magnetic Helicity: Evolution and Signatures'',
2016, in preparation.

\bibitem{bkt15}
A. Brandenburg, T. Kahniashvili, A. Tevzadze,
``Absence of the Inverse Cascade for Inflation-Generated Magnetic Fields",
2016, in preparation.

%\cite{Brandenburg:2016odr}
\bibitem{Brandenburg:2016odr}
  A.~Brandenburg and T.~Kahniashvili,
  %``Classes of hydrodynamic and magnetohydrodynamic turbulent decay,''
  arXiv:1607.01360 [physics.flu-dyn].
  %%CITATION = ARXIV:1607.01360;%%


\bibitem{fermi} E. Fermi,
"On the Origin of the Cosmic Radiation",
Phys. Rev. {\bf 75}, 1169 (1949).

%%%%%%%%%%%%%%%%%%%%%%%%%%%%%%%%%%%%%%

%\bibitem{phase}
\bibitem{phase}
E. R. Harrison,
%``Generation of magnetic fields in the radiation ERA'',
Mon. R. Astron. Soc. {\bf 147}, 279 (1970).

\bibitem{t1}
T. Vachaspati,
%``Magnetic fields from cosmological phase transitions'',
Phys.\ Lett.\  B {\bf 265}, 258 (1991).


\bibitem{t3}
J. M. Cornwall,
%``Speculations on primordial magnetic helicity'',
Phys. Rev. {\bf D 56}, 6146, (1997).

\bibitem{t4}
G. Sigl, A. V. Olinto and K. Jedamzik,
%``Primordial magnetic fields from cosmological first order phase transitions'',
Phys. Rev. D {\bf 55}, 4582 (1997).

\bibitem{t5}
M. Joyce and M. E. Shaposhnikov,
%``Primordial Magnetic Fields, Right Electrons, and the Abelian Anomaly'',
Phys. Rev. Lett. {\bf 79}, 1193 (1997).

\bibitem{t6}
M. Hindmarsh and A. Everett,
%``Magnetic fields from phase transitions'',
Phys. Rev. D {\bf 58}, 103505 (1998).

\bibitem{t7}
K. Enqvist,
%``Primordial Magnetic Fields'',
Int. J. Mod. Phys. {\bf D 7}, 331 (1998).

%\cite{Grasso:1997nx}
\bibitem{Grasso:1997nx}
  D.~Grasso and A.~Riotto,
%  ``On the nature of the magnetic fields generated during the electroweak phase transition,''
  Phys.\ Lett.\ B {\bf 418}, 258 (1998). %[hep-ph/9707265].
  %%CITATION = HEP-PH/9707265;%%
  %32 citations counted in INSPIRE as of 10 Nov 2014

\bibitem{t8}
J. Ahonen and K. Enqvist,
%``Magnetic field generation in first order phase transition bubble collisions'',
Phys. Rev. {\bf D 57}, 664 (1998).

\bibitem{t9}
M. Giovannini,
%``Primordial hypermagnetic knots'',
Phys. Rev. {\bf D 61}, 063004 (2000).

\bibitem{t10}
T. Vachaspati,
%``Estimate of the Primordial Magnetic Field Helicity'',
Phys. Rev. Lett. {\bf 87}, 251302 (2001).

\bibitem{t11}
A. D. Dolgov and D. Grasso,
%``Generation of Magnetic Fields and Gravitational Waves at Neutrino Decoupling'',
Phys. Rev. Lett. {\bf 88}, 011301 (2002).

\bibitem{t12}
D. Grasso and A. Dolgov, Nucl. Phys. Proc. Suppl. {\bf 110}, 189 (2002).

\bibitem{t13}
D. Boyanovsky, H. J.de Vega, and M. Simionato,
%``Magnetic field generation from nonequilibrium phase transitions'',
Phys. Rev. {\bf D 67}, 023502 (2003).

\bibitem{t14}
D. Boyanovsky, H. J. de Vega and M. Simionato,
%``Large scale magnetogenesis from a nonequilibrium phase transition in the radiation dominated era'',
Phys. Rev. {\bf D 67}, 123505 (2003).

\bibitem{t15}
A. Diaz-Gil, J. Garcia-Bellido, M. Garcia Perez, and A. Gonzalez-Arroyo,
%``Magnetic Field Production during Preheating at the Electroweak Scale'',
Phys. Rev. Lett. {\bf 100}, 241301 (2008).

\bibitem{t16}
T. Stevens, M.~B. Johnson, L.~S. Kisslinger, E.~M. Henley, W.-Y. P. Hwang,
and M. Burkardt,
%``Role of charged gauge fields in generating magnetic seed fields in bubble collisions during the cosmological electroweak phase transition'',
Phys. Rev. {\bf D 77}, 023501 (2008).

\bibitem{Quashnock:1988vs}
  J.~M.~Quashnock, A.~Loeb and D.~N.~Spergel,
%  ``Magnetic Field Generation During the Cosmological QCD Phase Transition,''
  Astrophys.\ J.\  {\bf 344}, L49 (1989).
  %%CITATION = ASJOA,344,L49;%%
  %119 citations counted in INSPIRE as of 14 Nov 2014

\bibitem{t17}
E.~M.~Henley, M.~B.~Johnson, and L.~S.~Kisslinger,
%``Electroweak phase transition nucleation with the MSSM and electromagnetic field creation'',
Phys.\ Rev.\  D {\bf 81}, 085035 (2010).

\bibitem{t18}
F.~R.~Urban and A.~R.~Zhitnitsky,
%``Large-scale magnetic fields, dark energy, and QCD'',
Phys.\ Rev.\  D {\bf 82}, 043524 (2010).

%\cite{Stevens:2012zz}
\bibitem{Stevens:2012zz}
  T.~Stevens, M.~B.~Johnson, L.~S.~Kisslinger and E.~M.~Henley,
%  ``Non-Abelian Higgs model of magnetic field generation during a cosmological first-order electroweak phase transition,''
  Phys.\ Rev.\ D {\bf 85}, 063003 (2012).
  %%CITATION = PHRVA,D85,063003;%%
  %1 citations counted in INSPIRE as of 10 Nov 2014

\bibitem{hogan} C.\ J.\ Hogan,
%"Magnetohydrodynamic effects of a first-order cosmological phase transition",
Phys.\ Rev.\ Lett.\  {\bf 51}, 1488 (1983).

\bibitem{causal} R.\ Durrer and C.\ Caprini,
%``Primordial magnetic fields and causality,''
JCAP {\bf 0311}, 010 (2003).

\bibitem{dav} P. A. Davidson, {\it  Turbulence} (Oxford: Oxford University Press, 2004).

%\cite{Caprini:2001nb}
\bibitem{Caprini:2001nb}
  C.~Caprini and R.~Durrer,
 % "Gravitational wave production: A Strong constraint on primordial magnetic fields,''
  Phys.\ Rev.\ D {\bf 65}, 023517 (2001).
  %[astro-ph/0106244].
  %%CITATION = ASTRO-PH/0106244;%%
  %103 citations counted in INSPIRE as of 09 Nov 2014

%%%%%%%%%%%%%%%%%%%%%%%%%%%%%%%%%%%%%%%

\bibitem{Cornwall:1997ms}
    J.~M.~Cornwall,
%    ``Speculations on primordial magnetic helicity,''
    Phys.\ Rev.\  D {\bf 56}, 6146 (1997).
    %[arXiv:hep-th/9704022].
    %%CITATION = PHRVA,D56,6146;%%

\bibitem{Giovannini:1997eg}
    M.~Giovannini and M.~E.~Shaposhnikov,
%    ``Primordial hypermagnetic fields and triangle anomaly,''
    Phys.\ Rev.\  D {\bf 57}, 2186 (1998).
    %[arXiv:hep-ph/9710234].
    %%CITATION = PHRVA,D57,2186;%%

\bibitem{Field:1998hi}
    G.~B.~Field and S.~M.~Carroll,
%    ``Cosmological magnetic fields from primordial helicity,''
    Phys.\ Rev.\  D {\bf 62}, 103008 (2000).
    %[arXiv:astro-ph/9811206].
    %%CITATION = PHRVA,D62,103008;%%

\bibitem{Vachaspati:2001nb}
    T.~Vachaspati,
%    ``Estimate of the primordial magnetic field helicity,''
    Phys.\ Rev.\ Lett.\  {\bf 87}, 251302 (2001).
    %[arXiv:astro-ph/0101261].
    %%CITATION = PRLTA,87,251302;%%

\bibitem{Tashiro:2012mf}
    H.~Tashiro, T.~Vachaspati and A.~Vilenkin,
%    ``Chiral Effects and Cosmic Magnetic Fields,''
    Phys.\ Rev.\ D {\bf 86}, 105033 (2012).
    %[arXiv:1206.5549 [astro-ph.CO]].
    %%CITATION = ARXIV:1206.5549;%%
    %1 citations counted in INSPIRE as of 09 May 2013

\bibitem{Sigl:2002kt}
    G.~Sigl,
%    ``Cosmological Magnetic Fields from Primordial Helical Seeds,''
    Phys.\ Rev.\  D {\bf 66}, 123002 (2002).
    %[arXiv:astro-ph/0202424].
    %%CITATION = PHRVA,D66,123002;%%

\bibitem{Subramanian:2004uf}
    K.~Subramanian and A.~Brandenburg,
%    ``Nonlinear current helicity fluxes in turbulent dynamos and alpha quenching,''
    Phys.\ Rev.\ Lett.\  {\bf 93}, 205001 (2004).
    %[arXiv:astro-ph/0408020].
    %%CITATION = PRLTA,93,205001;%%

\bibitem{Campanelli:2005ye}
    L.~Campanelli and M.~Giannotti,
%    ``Magnetic helicity generation from the cosmic axion field,''
    Phys.\ Rev.\  D {\bf 72}, 123001 (2005).
    %[arXiv:astro-ph/0508653].
    %%CITATION = PHRVA,D72,123001;%%

\bibitem{Semikoz:2004rr}
    V.~B.~Semikoz and D.~D.~Sokoloff,
%    ``Magnetic helicity and cosmological magnetic field,''
    Astron.\  Astrophys.\  {\bf 433}, L53 (2005);
    %[astro-ph/0411496];
    ``Large-scale cosmological magnetic fields and magnetic helicity,''
    Int.\ J.\ Mod.\ Phys.\  D {\bf 14}, 1839 (2005).
    %%CITATION = IMPAE,D14,1839;%%

\bibitem{DiazGil:2007dy}
    A.~Diaz-Gil, J.~Garcia-Bellido, M.~Garcia Perez and A.~Gonzalez-Arroyo,
%    ``Magnetic field production during preheating at the electroweak scale,''
    Phys.\ Rev.\ Lett.\  {\bf 100}, 241301 (2008).
    %[arXiv:0712.4263 [hep-ph]].
    %%CITATION = PRLTA,100,241301;%%

\bibitem{Campanelli:2008kh}
    L.~Campanelli,
%    ``Helical Magnetic Fields from Inflation,''
    Int.\ J.\ Mod.\ Phys.\ D {\bf 18}, 1395 (2009).
    %[arXiv:0805.0575 [astro-ph]].
    %%CITATION = ARXIV:0805.0575;%%

\bibitem{Campanelli:2013mea}
    L.~Campanelli,
%    ``Origin of Cosmic Magnetic Fields,''
    Phys.\ Rev.\ Lett.\  {\bf 111}, no. 6, 061301 (2013).
    %[arXiv:1304.6534 [astro-ph.CO]].
    %%CITATION = ARXIV:1304.6534;%%

\bibitem{Jain:2012jy}
    R.~K.~Jain, R.~Durrer and L.~Hollenstein,
%    ``Generation of helical magnetic fields from inflation,''
    J.\ Phys.\ Conf.\ Ser.\  {\bf 484}, 012062 (2014).
    %[arXiv:1204.2409 [astro-ph.CO]].
    %%CITATION = ARXIV:1204.2409;%%
    %4 citations counted in INSPIRE as of 09 Sep 2014


\bibitem{Long:2013tha}
    A.~J.~Long, E.~Sabancilar and T.~Vachaspati,
%    ``Leptogenesis and Primordial Magnetic Fields,''
    JCAP {\bf 1402}, 036 (2014).
    %[arXiv:1309.2315 [astro-ph.CO]].
    %%CITATION = ARXIV:1309.2315;%%
    %7 citations counted in INSPIRE as of 09 Sep 2014

\bibitem{Tashiro:2013ita}
  H.~Tashiro, W.~Chen, F.~Ferrer and T.~Vachaspati,
%  ``Search for CP Violating Signature of Intergalactic Magnetic Helicity in the Gamma Ray Sky,''
  Mon. Not. Roy. Astron. Soc.
  {\bf 445},  L41 (2014). %[arXiv:1310.4826 [astro-ph.CO]].
  %%CITATION = ARXIV:1310.4826;%%
  %5 citations counted in INSPIRE as of 13 Nov 2014

  \bibitem{Tashiro:2014gfa}
  H.~Tashiro and T.~Vachaspati,
%  ``Parity-odd correlators of diffuse gamma rays and intergalactic magnetic fields,''
Mon.\ Not.\ Roy.\ Astron.\ Soc.\  {\bf 448}, 299 (2015).
  %arXiv:1409.3627 [astro-ph.CO].
  %%CITATION = ARXIV:1409.3627;%%
%%%%%%%%%%%%%%%%%%%%%%%%%%%%%%%%%%%%%%%%%%%%%%%%%%%%%%%%%%%%%%%%%


%%%%INFLATION

\bibitem{inflation}
M.~S.~Turner and L.~M.~Widrow,
%"Inflation Produced, Large Scale Magnetic Fields,''
    Phys.\ Rev.\ D {\bf 37}, 2743 (1988).

\bibitem{ratra} B.\ Ratra,
%"Cosmological 'seed' magnetic field from inflation",
Astrophys.\ J.\  {\bf 391}, L1 (1992).

\bibitem{i1}
K.\ Bamba and J.\ Yokoyama,
%"Large-scale magnetic fields from dilaton inflation in noncommutative spacetime,''
Phys.\ Rev.\  D {\bf 69}, 043507 (2004).

\bibitem{i2}
L.\ Motta and R.\ R.\ Caldwell,
%  ``Non-Gaussian features of primordial magnetic fields in  power-law inflation,''
  Phys.\ Rev.\ D {\bf 85}, 103532 (2012).

\bibitem{i3}
R.\ R. Caldwell, L.\ Motta and M.\ Kamionkowski,
%``Correlation of inflation-produced magnetic fields with scalar fluctuations,''
Phys.\ Rev.\ D  {\bf 84}, 123525 (2011).

\bibitem{i4}
 D.\ Lemoine and M.\ Lemoine,
%   ``Primordial magnetic fields in string cosmology,''
   Phys.\ Rev.\ D {\bf 52}, 1955 (1995).  %%CITATION = PHRVA,D52,1955;%%

\bibitem{i5}
M.\ Gasperini, M.\ Giovannini and G.\ Veneziano,
%    ``Primordial magnetic fields from string cosmology,''
    Phys.\ Rev.\ Lett.\  {\bf 75}, 3796 (1995).

\bibitem{i6}
G.\ Lambiase, S.\ Mohanty and G.\ Scarpetta,
%``Magnetic field amplification in f(R) theories of gravity,''
    JCAP {\bf 0807}, 019 (2008).

\bibitem{i7}
L.\ Campanelli, P.\ Cea, G.\ L.\ Fogli and L.\ Tedesco,
%  ``Inflation-Produced Magnetic Fields in Nonlinear Electrodynamics,''
  Phys.\ Rev.\ D {\bf 77}, 043001 (2008).

\bibitem{i8}
L.\ Campanelli and P.\ Cea,
%  ``Maxwell-Kosteleck\'y Electromagnetism and Cosmic Magnetization,''
  Phys.\ Lett.\ B {\bf 675}, 155 (2009).

\bibitem{i9}
K.\ E.\ Kunze,
%  ``Large scale magnetic fields from gravitationally coupled electrodynamics,''
  Phys.\ Rev.\ D {\bf 81}, 043526 (2010).

\bibitem{Bamba:2012mi}
  K.~Bamba, C.~Q.~Geng and L.~W.~Luo,
%  ``Generation of large-scale magnetic fields from inflation in teleparallelism,''
  JCAP {\bf 1210}, 058 (2012). %[arXiv:1208.0665 [astro-ph.CO]].
  %%CITATION = ARXIV:1208.0665;%%
  %9 citations counted in INSPIRE as of 09 Nov 2014

%\cite{Jimenez:2011uia}
\bibitem{Jimenez:2011uia}
  J.~B.~Jim�nez and A.~L.~Maroto,
%  ``Dark energy and cosmic magnetic fields: electromagnetic relics from inflation,''
  Springer Proc.\ Phys.\  {\bf 137}, 215 (2011).
  %%CITATION = SPPPE,137,215;%%

%\cite{Membiela:2012ju}
\bibitem{Membiela:2012ju}
  F.~A.~Membiela and M.~Bellini,
%  ``Seminal magnetic fields from Inflato-electromagnetic Inflation,''
  Eur.\ Phys.\ J.\ C {\bf 72}, 2181 (2012).
  %[arXiv:1206.1873 [gr-qc]].
  %%CITATION = ARXIV:1206.1873;%%
  %2 citations counted in INSPIRE as of 09 Nov 2014

%\cite{Fujita:2012rb}
\bibitem{Fujita:2012rb}
  T.~Fujita and S.~Mukohyama,
%  ``Universal upper limit on inflation energy scale from cosmic magnetic field,''
  JCAP {\bf 1210}, 034 (2012). %
  %  [arXiv:1205.5031 [astro-ph.CO]].
  %%CITATION = ARXIV:1205.5031;%%
  %33 citations counted in INSPIRE as of 09 Nov 2014

%\cite{Motta:2012rn}
\bibitem{Motta:2012rn}
  L.~Motta and R.~R.~Caldwell,
%  ``Non-Gaussian features of primordial magnetic fields in power-law inflation,''
  Phys.\ Rev.\ D {\bf 85}, 103532 (2012).
  %[arXiv:1203.1033 [astro-ph.CO]].
  %%CITATION = ARXIV:1203.1033;%%
  %19 citations counted in INSPIRE as of 09 Nov 2014

%\cite{Bonvin:2011dt}
\bibitem{Bonvin:2011dt}
  C.~Bonvin, C.~Caprini and R.~Durrer,
%  ``Magnetic fields from inflation: the transition to the radiation era,''
  Phys.\ Rev.\ D {\bf 86}, 023519 (2012).
  %[arXiv:1112.3901 [astro-ph.CO]].
  %%CITATION = ARXIV:1112.3901;%%
  %21 citations counted in INSPIRE as of 09 Nov 2014
%\cite{Byrnes:2011aa}

\bibitem{Byrnes:2011aa}
  C.~T.~Byrnes, L.~Hollenstein, R.~K.~Jain and F.~R.~Urban,
%  ``Resonant magnetic fields from inflation,''
  JCAP {\bf 1203}, 009 (2012). %[arXiv:1111.2030 [astro-ph.CO]].
  %%CITATION = ARXIV:1111.2030;%%
  %17 citations counted in INSPIRE as of 09 Nov 2014

%\cite{Caldwell:2011ra}
\bibitem{Caldwell:2011ra}
  R.~R.~Caldwell, L.~Motta and M.~Kamionkowski,
%  ``Correlation of inflation-produced magnetic fields with scalar fluctuations,''
  Phys.\ Rev.\ D {\bf 84}, 123525 (2011).
  %  [arXiv:1109.4415 [astro-ph.CO]].
  %%CITATION = ARXIV:1109.4415;%%
  %33 citations counted in INSPIRE as of 09 Nov 2014

%\cite{Bamba:2011si}
\bibitem{Bamba:2011si}
  K.~Bamba, C.~Q.~Geng, S.~H.~Ho and W.~F.~Kao,
%  ``Large-scale magnetic fields from inflation due to a $CPT$-even Chern-Simons-like term with Kalb-Ramond and scalar fields,''
  Eur.\ Phys.\ J.\ C {\bf 72}, 1978 (2012). %[arXiv:1108.0151 [astro-ph.CO]].
  %%CITATION = ARXIV:1108.0151;%%
  %3 citations counted in INSPIRE as of 09 Nov 2014

%\cite{Martin:2007ue}
\bibitem{Martin:2007ue}
  J.~Martin and J.~Yokoyama,
 % ``Generation of Large-Scale Magnetic Fields in Single-Field Inflation,''
  JCAP {\bf 0801}, 025 (2008). %[arXiv:0711.4307 [astro-ph]].
  %%CITATION = ARXIV:0711.4307;%%
  %114 citations counted in INSPIRE as of 09 Nov 2014

%\cite{Jimenez:2010hu}
\bibitem{Jimenez:2010hu}
  J.~Beltran Jimenez and A.~L.~Maroto,
%  ``Cosmological magnetic fields from inflation in extended electromagnetism,''
  Phys.\ Rev.\ D {\bf 83}, 023514 (2011).  %[arXiv:1010.3960 [astro-ph.CO]].
  %%CITATION = ARXIV:1010.3960;%%
  %12 citations counted in INSPIRE as of 09 Nov 2014

%\cite{Durrer:2010mq}
\bibitem{Durrer:2010mq}
  R.~Durrer, L.~Hollenstein and R.~K.~Jain,
%  ``Can slow roll inflation induce relevant helical magnetic fields?,''
  JCAP {\bf 1103}, 037 (2011). %[arXiv:1005.5322 [astro-ph.CO]].
  %%CITATION = ARXIV:1005.5322;%%
  %41 citations counted in INSPIRE as of 09 Nov 2014

%\cite{Das:2010ywa}
\bibitem{Das:2010ywa}
  M.~Das and S.~Mohanty,
%  ``Magnetic Field Generation in Higgs Inflation Model,''
  Int.\ J.\ Mod.\ Phys.\ A {\bf 27}, 1250040 (2012). %[arXiv:1004.1927 [astro-ph.CO]].
  %%CITATION = ARXIV:1004.1927;%%
  %1 citations counted in INSPIRE as of 09 Nov 2014

%\cite{Membiela:2010rv}
\bibitem{Membiela:2010rv}
  F.~A.~Membiela and M.~Bellini,
%  ``Coupled inflaton and electromagnetic fields from Gravitoelectromagnetic Inflation with Lorentz and Feynman gauges,''
  JCAP {\bf 1010}, 001 (2010). %[arXiv:1003.4175 [astro-ph.CO]].
  %%CITATION = ARXIV:1003.4175;%%

\bibitem{Demozzi:2009fu}
V.\ Demozzi, V.\ Mukhanov and H.\ Rubinstein,
%``Magnetic fields from inflation?,'
JCAP {\bf 0908}, 025 (2009).

%\cite{Kanno:2009ei}
\bibitem{Kanno:2009ei}
  S.~Kanno, J.~Soda and M.~a.~Watanabe,
%  ``Cosmological Magnetic Fields from Inflation and Backreaction,''
  JCAP {\bf 0912}, 009 (2009). %[arXiv:0908.3509 [astro-ph.CO]].
  %%CITATION = ARXIV:0908.3509;%%
  %83 citations counted in INSPIRE as of 09 Nov 2014

%\cite{Urban:2011bu}
\bibitem{Urban:2011bu}
  F.~R.~Urban,
%  ``On inflating magnetic fields, and the backreactions thereof,''
  JCAP {\bf 1112}, 012 (2011). %[arXiv:1111.1006 [astro-ph.CO]].
  %%CITATION = ARXIV:1111.1006;%%
  %9 citations counted in INSPIRE as of 09 Nov 2014

\bibitem{b22}
V.\ Demozzi and C.\ Ringeval,
%"Reheating constraints in inflationary magnetogenesis,''
  JCAP {\bf 1205}, 009 (2012).

\bibitem{Tasinato:2014fia}
  G.~Tasinato,
%  ``A scenario for inflationary magnetogenesis without strong coupling problem,''
  JCAP {\bf 1503}, 040 (2015).
  %arXiv:1411.2803 [hep-th].
  %%CITATION = ARXIV:1411.2803;%%

%%%%%%%%%%%%%%%%%%%%%%%%%%%%%%%%%%%%%%%%%%%%%%%%%%%%%%%%%%%%%%%
%%%%%%%%%%%%%%%%%%%%%%%%%%%%%%%%%%%%%%%%%%%%%%%%%%%%%%%%%%%%%%%%%

%\cite{Brandenburg:2013vya}
\bibitem{Brandenburg:2013vya}
  A.~Brandenburg and A.~Lazarian,
%  ``Astrophysical hydromagnetic turbulence,''
  Space Sci.\ Rev.\  {\bf 178}, 163 (2013).
  %arXiv:1307.5496 [astro-ph.SR].
  %%CITATION = ARXIV:1307.5496;%%


%\cite{Colgate:2000gb}
\bibitem{Colgate}
  S.~A.~Colgate, H.~Li and V.~Pariev,
%  ``The origin of the magnetic fields of the universe: the plasma astrophysics of the free energy of the universe,''
  astro-ph/0012484.
  %%CITATION = ASTRO-PH/0012484;%%

%\cite{Xu:2009if}
\bibitem{Xu:2009if}
  H.~Xu, H.~Li, D.~C.~Collins, S.~Li and M.~L.~Norman,
%  ``Turbulence and Dynamo in Galaxy Cluster Medium: Implications on the Origin of Cluster Magnetic Fields,''
  Astrophys.\ J.\  {\bf 698}, L14 (2009).   %[arXiv:0905.2196 [astro-ph.CO]].  %%CITATION = ARXIV:0905.2196;%%

\bibitem{nature2012} G. Gregori,
A. {Ravasio}, {\it et al.}
%C. D.  {Murphy}, K. {Schaar}, A. {Baird}, A. R. {Bell}, A.  {Benuzzi-Mounaix}, R. {Bingham}, C. {Constantin},  R. P. {Drake}, M. {Edwards}, E. T. {Everson}, C. D. {Gregory}, Y. {Kuramitsu}, W. {Lau}, J.  {Mithen}, C. {Niemann}, H. S. {Park}, B. A. {Remington}, B.  {Reville}, A. P. L.  {Robinson}, D. D. {Ryutov},  Y. {Sakawa}, S. {Yang}, N. C.  {Woolsey}, M. {Koenig}, F. {Miniati},
%"Generation of scaled protogalactic seed magnetic fields in laser-produced shock waves",
Nature, {\bf 481}, 480 (2012).

\bibitem{Alves:2011pp}
  E.~P.~Alves, T.~Grismayer, S.~F.~Martins, F.~Fiuza, R.~A.~Fonseca and L.~O.~Silva,
 % ``Large-scale magnetic field generation via the kinetic Kelvin-Helmholtz instability in unmagnetized scenarios,''
  Astrophys. J. {\bf 746}. L14 (2012).
  %%CITATION = ARXIV:1107.6037;%%

\bibitem{amplification} J.~Schober, D.~Schleicher, C.~Federrath, R.~Klessen and R.~Banerjee,
%``Magnetic Field Amplification by Small-Scale Dynamo Action: Dependence on Turbulence Models and Reynolds and Prandtl Numbers,''
  Phys. Rev. E. {\bf 85}, 026303 (2012).

  \bibitem{am1}
  S.~Sur, C.~Federrath, D.~R.~G.~Schleicher, R.~Banerjee and R.~S.~Klessen,
%  ``Magnetic field amplification during gravitational collapse - Influence of initial conditions on dynamo evolution and saturation,''
Mon. Roy. Not. Astron. Sci. {\bf 423} 3148 (2012).
  %arXiv:1202.3206 [astro-ph.SR].
  %arXiv:1109.4571 [astro-ph.CO];

%%%%%%%%%%%%%%%%%%%%%%%%%%%%%%%%%
\bibitem{beo96} A.~Brandenburg, K.~Enqvist and P.~Olesen,
% ``Large-scale magnetic fields from hydromagnetic turbulence in the very early universe,''
  Phys.\ Rev.\  D {\bf 54}, 1291 (1996).

\bibitem{axel-mode}
A. Brandenburg and A. Nordlund,
%"Astrophysical turbulence modeling",
Rep. Prog. Phys. {\bf 74}, 046901 (2011).


\bibitem{robi}
R. Banarjee,
%"Evolution of Primordial Magnetic Field",
Astron. Nach. {\bf 334}, 537 (2013).


\bibitem{Subramanian:1997gi}
  K.~Subramanian and J.~D.~Barrow,
%  ``Magnetohydrodynamics in the early universe and the damping of noninear Alfven waves,''
  Phys.\ Rev.\ D {\bf 58}, 083502 (1998). %[astro-ph/9712083].
  %%CITATION = ASTRO-PH/9712083;%%
  %154 citations counted in INSPIRE as of 14 Nov 2014



%\cite{Dimopoulos:1996nq}
\bibitem{Dimopoulos:1996nq}
  K.~Dimopoulos and A.~C.~Davis,
%  ``On the evolution of primordial magnetic fields,''
  Phys.\ Lett.\ B {\bf 390}, 87 (1997).
  %[astro-ph/9610013].
  %%CITATION = ASTRO-PH/9610013;%%
  %33 citations counted in INSPIRE as of 14 Nov 2014

\bibitem{Wagstaff:2013yna}
  J.~M.~Wagstaff, R.~Banerjee, D.~Schleicher and G.~Sigl,
%  ``Magnetic field amplification by the small-scale dynamo in the early Universe,''
  Phys.\ Rev.\ D {\bf 89}, 103001 (2014). %[arXiv:1304.4723 [astro-ph.CO]].
  %%CITATION = ARXIV:1304.4723;%%
  %8 citations counted in INSPIRE as of 14 Nov 2014

\bibitem{banerjee} R.~Banerjee and K.~Jedamzik,
%  ``The Evolution of Cosmic Magnetic Fields: From the Very Early Universe, to Recombination, to the Present,''
  Phys.\ Rev.\  D {\bf 70}, 123003 (2004).

\bibitem{kulsrud}
T. Tajima, S. Cable, K. Shibata, R. M. Kulsrud,
% "On the origin of cosmological magnetic fields"
Astrophys. J. {\bf 390 } 309 (1992).

%

\bibitem{db} A.~Brandenburg and W.~Dobler,
%  ``Hydromagnetic turbulence in computer simulations,''
  Comput.\ Phys.\ Commun.\  {\bf 147}, 471 (2002).

\bibitem{PFL}
A. Pouquet, U. Frisch and J. L\'eorat,
J. Fluid Mech. {\bf 77}. 321 (1976).

\bibitem{Jedamzik:2010cy}
  K.~Jedamzik and G.~Sigl,
%  ``The Evolution of the Large-Scale Tail of Primordial Magnetic Fields,''
  Phys.\ Rev.\ D {\bf 83}, 103005 (2011).
  %[arXiv:1012.4794 [astro-ph.CO]].
  %%CITATION = ARXIV:1012.4794;%%
  %19 citations counted in INSPIRE as of 15 Nov 2014

%\cite{Saveliev:2012ea}
\bibitem{Saveliev:2012ea}
  A.~Saveliev, K.~Jedamzik and G.~Sigl,
%  ``Time Evolution of the Large-Scale Tail of Non-Helical Primordial Magnetic Fields with Back-Reaction of the Turbulent Medium,''
  Phys.\ Rev.\ D {\bf 86}, 103010 (2012).
  %[arXiv:1208.0444 [astro-ph.CO]].
  %%CITATION = ARXIV:1208.0444;%%
  %6 citations counted in INSPIRE as of 15 Nov 2014

%%%%%%%%%%%%%%%%%%%%%



\bibitem{Berera:2014hca}
  A.~Berera and M.~Linkmann,
  ``Magnetic helicity and the evolution of decaying magnetohydrodynamic turbulence,''
  Phys.\ Rev.\ E {\bf 90}, no. 4, 041003 (2014). %[arXiv:1405.6756 [physics.flu-dyn]].
  %%CITATION = ARXIV:1405.6756;%%
  %1 citations counted in INSPIRE as of 14 Nov 2014



\bibitem{Zrake:2014mta}
  J.~Zrake,
%  ``Inverse cascade of non-helical magnetic turbulence in a relativistic fluid,''
  Astrophys.\ J.\  {\bf 794}, no. 2, L26 (2014).
  %[arXiv:1407.5626 [astro-ph.HE]].
  %%CITATION = ARXIV:1407.5626;%%
  %1 citations counted in INSPIRE as of 14 Nov 2014

\bibitem{East:2015pea}
  W.~E.~East, J.~Zrake, Y.~Yuan and R.~D.~Blandford,
  %``Spontaneous decay of periodic magnetostatic equilibria,''
  arXiv:1503.04793 [astro-ph.HE].
  %%CITATION = ARXIV:1503.04793;%%

\bibitem{Galtier:2000ce}
  S.~Galtier, S.~V.~Nazarenko, A.~C.~Newell and A.~Pouquet,
%  ``A weak turbulence theory for incompressible mhd,''
  J.\ Plasma Phys.\  {\bf 63}, 447 (2000).
  %[astro-ph/0008148].
  %%CITATION = ASTRO-PH/0008148;%%
  %83 citations counted in INSPIRE as of 14 Nov 2014
%%%%%%%%%%%%%%%%%%%%%%%%%%%%%%%%%%%%%%%%%%%%%%%%%%%%%%%


\bibitem{bII} D. Biskamp, {\it  Nonlinear Magnetohydrodynamics} (Cambridge: Cambridge
Univ. Press) 2000.

\bibitem{b2} D. Biskamp and W.~C.~M\"uller, Phys.\ Rev.\ Lett., 83, 2195 (1999),
D. Biskamp and W.~C.~M\"uller, Phys.\ Plasma, 7, 4889 (2000).

\bibitem{Son:1998my}
  D.~T.~Son,
%  ``Magnetohydrodynamics of the early universe and the evolution of primordial magnetic fields,''
  Phys.\ Rev.\ D {\bf 59}, 063008 (1999). %[hep-ph/9803412].
  %%CITATION = HEP-PH/9803412;%%
  %90 citations counted in INSPIRE as of 14 Nov 2014

\bibitem{blackman}
E. Blackman,
%"Magnetic Helicity and Large Scale Magnetic Fields: A Primer",
Spa.\ Sci.\ Rev. {\bf 188}, 59 (2015). %arXiv:1402.0933 (2014).

\bibitem{Saveliev:2013uva}
  A.~Saveliev, K.~Jedamzik and G.~Sigl,
%  ``Evolution of Helical Cosmic Magnetic Fields as Predicted by Magnetohydrodynamic Closure Theory,''
  Phys.\ Rev.\ D {\bf 87}, no. 12, 123001 (2013).
  %[arXiv:1304.3621 [astro-ph.CO]].
  %%CITATION = ARXIV:1304.3621;%%
  %6 citations counted in INSPIRE as of 14 Nov 2014

\bibitem{campanelli2} L.~Campanelli,
%  ``Scaling laws in MHD turbulence,''
  Phys.\ Rev.\ {\bf D70}, 083009 (2004).

\bibitem{campanelli} L.~Campanelli,
%  ``Evolution of Magnetic Fields in Freely Decaying Magnetohydrodynamic Turbulence,''
  Phys.\ Rev.\ Lett.\  {\bf 98}, 251302 (2007).
%  [arXiv:0705.2308 [astro-ph]].



%\cite{Christensson:2002xu}
\bibitem{Christensson:2002xu}
  M.~Christensson, M.~Hindmarsh and A.~Brandenburg,
%  ``Scaling laws in decaying helical 3-D magnetohydrodynamic turbulence,''
  Astron.\ Nachr.\  {\bf 326}, 393 (2005). %[astro-ph/0209119].
  %%CITATION = ASTRO-PH/0209119;%%
  %37 citations counted in INSPIRE as of 14 Nov 2014

\bibitem{Christensson:2000sp}
  M.~Christensson, M.~Hindmarsh and A.~Brandenburg,
%  ``Inverse cascade in decaying 3-D magnetohydrodynamic turbulence,''
  Phys.\ Rev.\ E {\bf 64}, 056405 (2001).
  %[astro-ph/0011321].
  %%CITATION = ASTRO-PH/0011321;%%
  %50 citations counted in INSPIRE as of 14 Nov 2014

\bibitem{gal1}
K. Subramanian and A. Brandenburg,
%"Traces of large-scale dynamo action in the kinematic stage",
Mon. Not. Roy. Astron. Soc. {\bf 445}, 2930
(2014).

\bibitem{gal2}
L. Chamandy, A. Shukurov, K. Subramanian,
%"Magnetic spiral arms and galactic outflows"
Mon. Not. Roy. Astron. Soc. {\bf 446}, L6 (2015).

\bibitem{gal3}
E. A. Mikhailov,
%"Star formation and galactic dynamo model with helicity fluxes"
Astron. Lett. {\bf 40}, 398 (2014).

\bibitem{Wagstaff:2014fla}
  J.~M.~Wagstaff and R.~Banerjee,
 %``Extragalactic magnetic fields rule out electroweak phase transition magnetogenesis,''
  JCAP {\bf 1601}, 002 (2016).
  %arXiv:1409.4223 [astro-ph.CO].
  %%CITATION = ARXIV:1409.4223;%%
  %1 citations counted in INSPIRE as of 14 Nov 2014




\bibitem{mendis}
%Whipple, E. C., Northrop, T. G., and Mendis, D. A.,
E. C. Whipple, T. G. Northrop, and D. A. Mendis,
J. Geophys. Res. {\bf 90}, 7405 (1985).

\bibitem{Blackman}
Blackman, E. G.,
Spa. Sci. Rev. {\bf 188}, 59 (2015).
%Magnetic Helicity and Large Scale Magnetic Fields: A Primer


\end{thebibliography}
\end{document}